%% file: 06035.tex
\documentclass[10pt,twocolumn,letterpaper]{article}

\usepackage{cvpr}
\usepackage{times}
\usepackage{epsfig}
\usepackage{graphicx}
\usepackage{amsmath}
\usepackage{amssymb}
\usepackage{wrapfig}
\usepackage{balance}
\usepackage{longtable}
\input{include}
\usepackage[breaklinks=true,bookmarks=false]{hyperref}

\cvprfinalcopy % *** Uncomment this line for the final submission

\def\cvprPaperID{6035} % *** Enter the CVPR Paper ID here

\ifcvprfinal\fi

\begin{document}

%%%%%%%%% TITLE
\title{Neural Cages for Detail-Preserving 3D Deformations}
\author{
	Wang Yifan\textsuperscript{1}\hspace{1.0em}
	Noam Aigerman\textsuperscript{2}\hspace{1.0em}
	Vladimir G. Kim\textsuperscript{2}\hspace{1.0em}
	Siddhartha Chaudhuri \textsuperscript{2,3}\hspace{1.0em}
	Olga Sorkine-Hornung\textsuperscript{1}\hspace{1.0em}
	\\\\
	\textsuperscript{1}ETH Zurich \hspace{1.0em} \textsuperscript{2}Adobe Research
	\hspace{1.0em}  \textsuperscript{3}IIT Bombay
	\vspace{-8pt}
}

\maketitle
%\thispagestyle{empty}
%%%%%%%%% ABSTRACT
\begin{abstract}
We propose a novel learnable representation for detail-preserving shape deformation. The goal of our method is to warp a source shape to match the general structure of a target shape, while preserving the surface details of the source. Our method extends a traditional cage-based deformation technique, where the source shape is enclosed by a coarse control mesh termed \emph{cage}, and translations prescribed on the cage vertices are interpolated to any point on the source mesh via special weight functions. The use of this sparse cage scaffolding enables preserving surface details regardless of the shape's intricacy and topology. Our key contribution is a novel neural network architecture for predicting deformations by controlling the cage. We incorporate a differentiable cage-based deformation module in our architecture, and train our network end-to-end. Our method can be trained with common collections of 3D models in an unsupervised fashion, without any cage-specific annotations. We demonstrate the utility of our method for synthesizing shape variations and deformation transfer.
\end{abstract}

%%%%%%%%% BODY TEXT
\input{tex/introduction}
\input{tex/related}

\input{tex/method}
\input{tex/application}
\input{tex/evaluation}
\input{tex/conclusion}

\section*{Acknowledgments}
We thank Rana Hanocka and Dominic Jack for their extensive help. The robot model in Figures~\ref{fig:teaser} and \ref{fig:dt_composed} is from SketchFab, licensed under CC-BY. This work was supported in part by gifts from Adobe, Facebook and Snap.

{\small
\bibliographystyle{ieee_fullname}
\bibliography{egbib}
}
%\cleardoublepage
\input{06035-supp}
\end{document}

%% file: include.tex
\usepackage{color}
\usepackage[sort,nocompress]{cite}
\usepackage[font={footnotesize}]{caption}
\usepackage{subcaption}
\usepackage{xspace}
\usepackage{overpic}
\usepackage{tikz}
\usepackage[skins]{tcolorbox}
\usepackage{pgfplots}
\pgfplotsset{compat=newest}
\usetikzlibrary{spy,calc,shapes,arrows,positioning,pgfplots.groupplots}
\usepackage[]{paralist}

\usepackage{enumitem}
\usepackage{mathrsfs}
\usepackage{textcomp}
\usepackage{booktabs}
\usepackage{multirow}
\usepackage{makecell}
\usepackage{dsfont}

\usepackage{arydshln}

\newcommand{\finalcells}[2]{%
\begingroup\sbox0{\begin{minipage}{3cm}\raggedright#1\end{minipage}}%
\sbox2{\begin{minipage}{3cm}\raggedright#2\end{minipage}}%
\xdef\finalheight{\the\dimexpr\ht0+\dp0+\smallskipamount\relax}%
\xdef\finalheightB{\the\dimexpr\ht2+\dp2+\smallskipamount\relax}%
\ifdim\finalheightB>\finalheight
\global\let\finalheight\finalheightB
\fi\endgroup
\begin{minipage}[t][\finalheight][t]{3cm}\raggedright#1\end{minipage}&
\begin{minipage}[t][\finalheight][t]{3cm}\raggedright#2\end{minipage}}

\definecolor{turquoise}{cmyk}{0.65,0,0.1,0.1}
\definecolor{purple}{rgb}{0.65,0,0.65}
\definecolor{dark_green}{rgb}{0, 0.5, 0}
\definecolor{orange}{rgb}{0.8, 0.6, 0.2}
\definecolor{red}{rgb}{0.8, 0.2, 0.2}
\definecolor{brown}{rgb}{0.5, 0.16, 0.16}

\newcommand{\changed}[1]{{#1}}

\renewcommand{\vec}[1]{\ensuremath{\mathbf{#1}}}
\newcommand{\sh}{\mathcal{S}}
\newcommand{\cage}{\mathcal{C}}

\renewcommand{\to}{\rightarrow}
\newcommand{\cagev}{\vec{v}_j}
\newcommand{\shapev}{\vec{p}_i}

\newcommand{\net}{\mathcal{N}}
\newcommand{\B}{\mathcal{B}}
\newcommand{\nn}{\vec{n}}
\newcommand{\qq}{\vec{q}}
\newcommand{\pp}{\vec{p}}
\newcommand{\LL}{\mathcal{L}}
\makeatletter
\def\bblfootnote{\xdef\@thefnmark{}\@footnotetext}
\makeatother

\usepackage[skins]{tcolorbox} % for the teaser
\usepackage{shadowtext}
\shadowrgb{0,0,0}
\shadowoffset{0.5pt}

%% file: tex/introduction.tex
%!TEX root = ../06035.tex
\vspace{-4ex}
\section{Introduction}\label{sec:intro}
Deformation of 3D shapes is a ubiquitous task, arising in many vision and graphics applications. For instance, {\em deformation transfer}~\cite{sumner2004deformtransfer} aims to infer a deformation from a given pair of shapes and apply the same deformation to a novel target shape. As another example, a small dataset of shapes from a given category (\eg, chairs) can be augmented by {\em synthesizing variations}, where each variation deforms a randomly chosen shape to the proportions and morphology of another while preserving local detail~\cite{xu2010anisotropic,wang20193dn}. 

Deformation techniques usually need to simultaneously optimize at least two competing objectives. The first is alignment with the target, \eg, matching limb positions while deforming a human shape to another human in a different pose. The second objective is adhering to quality metrics, such as distortion minimization and preservation of local geometric features, such as the human's face. 
These two objectives are contradictory, since a perfect alignment of a deformed source shape to the target precludes preserving the original details of the source.

\input{tex/teaser}

Due to these conflicting objectives, optimization techniques~\cite{li08global} require parameter tuning to balance the two competing terms, and are heavily reliant on an inferred or manually supplied correspondence between the source and the target. These parameters vary based on the shape category, representation, and the level of dissimilarity between the source and the target.

To address these limitations, recent techniques train a \emph{neural network} to predict shape deformations. This is achieved by predicting new positions for all vertices of a template shape~\cite{liu2018meshVAE} or by implicitly representing the deformation as a mapping of all points in 3D, which is then used to map each vertex of a source shape~\cite{wang20193dn,groueix19cycleconsistentdeformation}. Examples of the results of some of these methods can be seen in Fig~\ref{fig:3Dsynthesis}, which demonstrates the limitations of such approaches: the predicted deformations corrupt features and exhibit distortion, especially in areas with thin structures, fine details or gross discrepancies between source and target. These artifacts are due to the inherent limitations of neural networks to capture, preserve, and generate high frequencies. 

In this paper, we circumvent the above issues via a classic geometry processing technique called \emph{cage-based deformation}~\cite{ju2005mean,lipman2008green,joshi2007harmonic}, abbreviated to \emph{CBD}. In CBD, the source shape is enclosed in a very coarse scaffold mesh called the \emph{cage} (Fig~\ref{fig:overview}). The deformation of the cage is transferred to the enclosed shape by interpolating the translations of the cage vertices. Fittingly, the interpolation schemes in these classic works are carefully designed to preserve details and minimize distortion.

Our main technical contribution is a novel neural architecture in which, given a source mesh, learnable parameters are optimized to predict both the positioning of the cage around the source shape, as well as the deformation of that cage, which drives the deformation of the enclosed shape in order to match a target shape. The source shape is deformed by deterministically interpolating the new positions of its surface points from those of the cage vertices, via a novel, differentiable, cage-based deformation layer. The pipeline is trained end-to-end on a collection of randomly chosen pairs of shapes from a training set.

The {\em first key advantage} of our method is that cages provide a much more natural space for predicting deformations: CBD is feature-preserving by construction, the degrees of freedom in deformation only depends on the number of vertices on the coarse cage. In short, our network makes a prediction in a low-dimensional space of highly regular deformations.

The {\em second key advantage} is that our method is not tied to a single source shape, nor to a single mesh topology. As the many examples in this paper demonstrate, the trained network can predict and deform cages for similar shapes not observed during training. The target shape can be crude and noisy, \eg, a shape acquired with cheap scanning hardware or reconstructed from an image. Furthermore, dense correspondences between the source and target shapes are not required in general, though they can help when the training set has very varied articulations. Thus the method can be trained on large datasets that are not co-registered and do not have consistently labeled landmarks.

We show the utility of our method in two main applications. We generate shape variations by deforming a 3D model using other shapes as well as images as targets. We also use our method to pose a human according to a target humanoid character, and, given a few sparse correspondences, perform deformation transfer and pose an arbitrary novel humanoid. See Figures~\ref{fig:teaser}, \ref{fig:svr_real}, \ref{fig:dt_composed} and \ref{fig:3Dsynthesis} for examples.

%% file: tex/teaser.tex
\begin{figure}[t!]
\includegraphics[width=\linewidth]{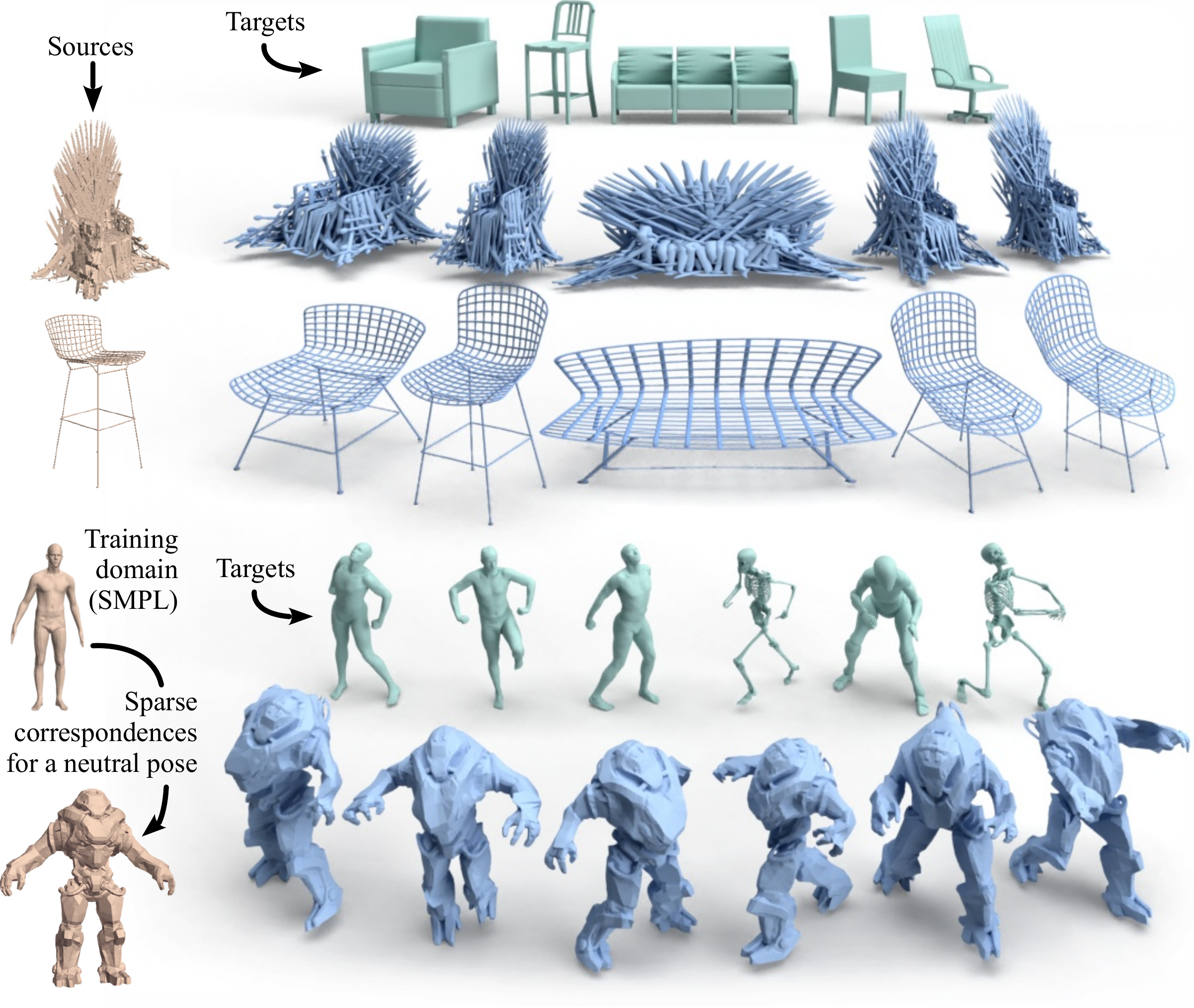}
\vspace{-.8cm}
\caption{Applications of our neural cage-based deformation method. {\em Top:} Complex source chairs (brown) deformed (blue) to match target chairs (green), while accurately preserving detail and style with non-homogeneous changes that adapt different regions differently. No correspondences are used at any stage. {\em Bottom:} A cage-based deformation network trained on many posed humans (SMPL) can transfer various poses of novel targets (SCAPE, skeleton, X-Bot, in green) to a very dissimilar robot of which only a single neutral pose is available. A few matching landmarks between the robot and a neutral SMPL human are required. Dense correspondences between SMPL humans are used only during training.}
\vspace{-0.1cm}
\label{fig:teaser}
\end{figure}

%% file: tex/related.tex
%!TEX root = ../06035.tex
\section{Related work}\label{sec:rw}

We now review prior work on learning deformations, traditional methods for shape deformation, and  applications.

\noindent \textbf{Learning 3D deformations.}
Many recent works in learning 3D geometry have focused on generative tasks, such as synthesis~\cite{groueix2018atlasnet,mescheder2019occupancy} and editing~\cite{zhu_siga18} of unstructured geometric data. These tasks are especially challenging if one desires high-fidelity content with intricate details. A common approach to producing intricate shapes is to deform an existing generic~\cite{wang2018pixel2mesh} or category-specific~\cite{groueix20183d} template. Early approaches represented deformations as a single vector of vertex
positions of a template~\cite{liu2018meshVAE}, which limited their output to shapes constructable by deforming the specific template, and also made the architecture sensitive to the template tessellation. An alternative is to predict a freeform deformation field over 3D voxels~\cite{jack2018learning,hanocka2018alignet,yumer2016learning}; however, this makes the deformation's resolution dependent on the voxel resolution, and thus has limited capability to adapt to a specific shape categories and source shapes.

Alternatively, some architectures learn to map a single point at a time, conditioned on some global descriptor of the target shape~\cite{groueix20183d}. These architectures can also work for novel sources by conditioning the deformation field on features of both source and target~\cite{groueix19cycleconsistentdeformation,wang20193dn}. Unfortunately, due to network capacity limits, these techniques struggle to represent intricate details and tend to blur high-frequency features.

\noindent \textbf{Traditional methods for mesh deformation.}
Research on detail-preserving deformations in the geometry processing community spans several decades and has contributed various formulations and optimization techniques~\cite{DeformationTutorial:2009}. These methods usually rely on a sparse set of control points whose transformations are interpolated to all remaining points of the shape; the challenge lies in defining this interpolation in a way that preserves details. This can be achieved by solving an optimization problem to reduce the distortion of the deformation such as ~\cite{ARAP_modeling:2007}. However, defining the output deformation as the solution to an intricate non-convex optimization problem significantly limits the ability of a network to learn this deformation space.

Instead, we use cage-based deformations as our representation, where the source shape is enclosed by a coarse \textit{cage} mesh, and all surface points are written as linear combinations of the cage vertices, i.e., generalized barycentric coordinates. Many designs have been proposed for these coordinate functions such that shape structure and details are preserved under interpolations~\cite{ju2005mean,lipman2008green,joshi2007harmonic,calderon2017bounding,TTB:CageR:CGF,Sacht:2015:NC:2816795.2818093,xian2012automatic}.

\noindent \textbf{Shape synthesis and deformation transfer.}
Automatically aligning a source shape to a target shape while preserving details is a common task, used to synthesize variations of shapes for amplification of stock datasets~\cite{Huang2015SRV} or for transferring a given deformation to a new model, targeting animation synthesis~\cite{sumner2004deformtransfer}. To infer the deformation, correspondence between the two shapes needs to be accounted for, either by explicitly inferring corresponding points~\cite{li08global,li2012temporally,Huang:2008}, or by implicitly conditioning the deformation fields on the latent code of the target shape~\cite{groueix19cycleconsistentdeformation,wang20193dn,hanocka2018alignet}. Our work builds upon the latter learning-based framework, but uses cages to parameterize the space of deformations.

Gao et al.~\cite{gao2018automatic} automate the deformation transfer for unpaired shapes using cycled generative adversarial networks, thus the trained method cannot be easily adapted for new shape targets. Some prior techniques focus on transferring and interpolating attributes between various latent spaces trained for shape generation~\cite{yin2019logan,gaosdmnet2019}. These generative models are not capable of fully preserving local geometric features, especially if the source is not pre-segmented into simpler primitives (as assumed by~\cite{gaosdmnet2019}). In general, such methods are only expected to perform well if the input shapes are relatively similar to those observed at training time.

%% file: tex/method.tex
%!TEX root = ../06035.tex
\section{Method}\label{sec:method}
We now detail our approach for learning cage-based deformations (CBD). We start with a brief overview of the principles of CBD, and then explain how we train a network to control these deformations from data. 
The implementation is available at \href{https://github.com/yifita/deep_cage}{https://github.com/yifita/deep\_cage}.

%\begin{figure*}\centering
%\def\svgwidth{0.95\linewidth}\input{overview.pdf_tex}
%\caption{An overview of our method. Given a pair of source and target shapes, our network jointly learns to predict an initial cage $ \cage_s $ around the source shape and the cage deformation which warps $ \sh_s $ to $ \sh_t $. The differentiable barycentric coordinates $ \phi $ interpolate the cage deformation to the embedded object, so that the deformed shape $ \sh_{s\to t} $ can be obtained using a simple linear combination. \yf{replace the part segmentation with non-segmented vases.}} \label{fig:overview}
%\end{figure*}
\begin{figure}[t!]
\includegraphics[width=\columnwidth]{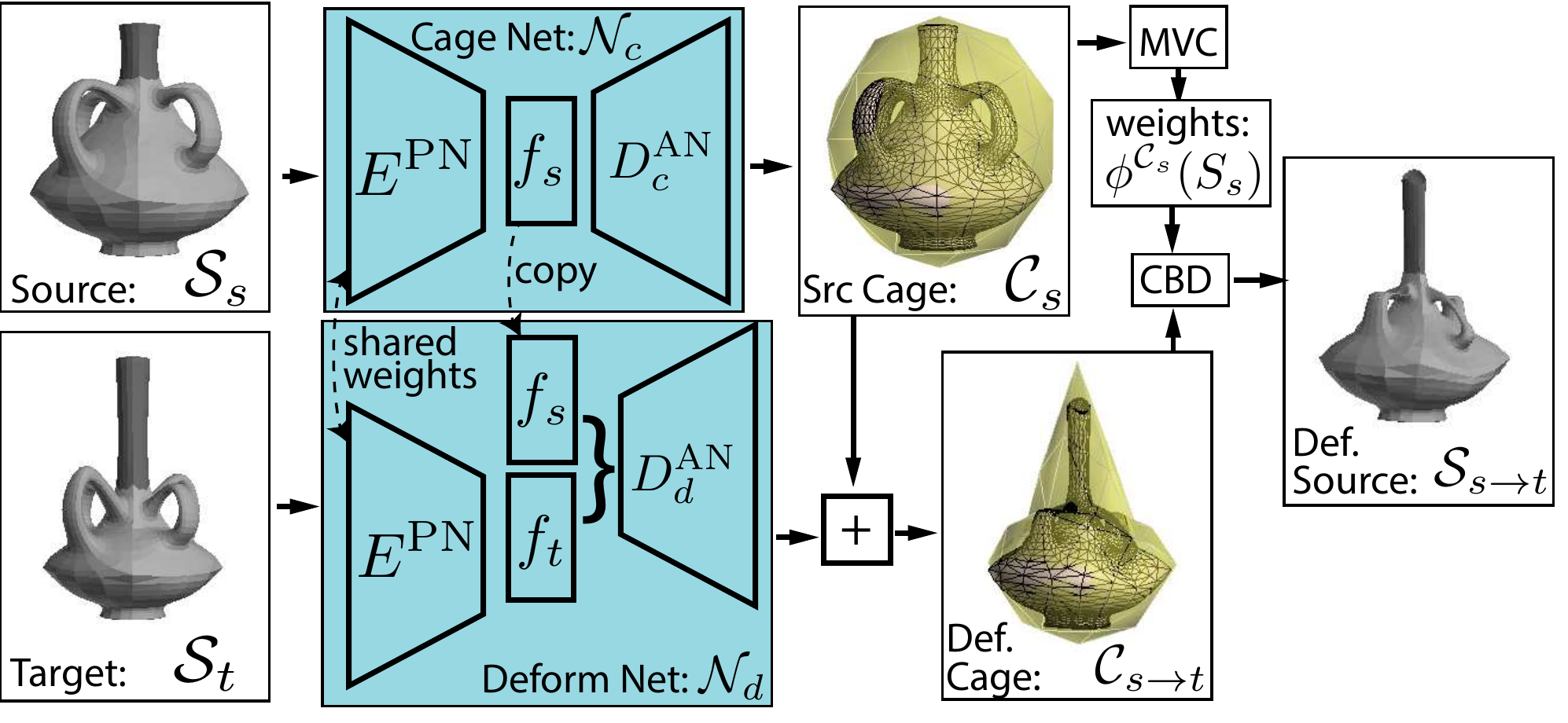}
\vspace{-0.7cm}
\caption{Overview. A source $\sh_s$ and a target $\sh_t$ are encoded by the same PointNet encoder $E^\text{PN}$ into latent codes $f_s$ and $f_t$, resp. An AtlasNet-style decoder $D^\text{AN}_c$ decodes $ f_s $ to a source cage $\cage_s$ in the cage module $\net_c$. Another decoder $D^\text{AN}_d$ creates the offset for $ \cage_s $ in the deformation module $\net_d$ from the concatenation of $ f_s $ and $ f_t $, yielding a deformed cage $ \cage_{s\to t} $. Given a source cage and shape, our novel MVC layer computes the mean value coordinates $\phi^{\cage_s}(\sh_s)$, which are used to produce a deformed source shape $\sh_{s\rightarrow t}$ from the cage deformation $\cage_{s\rightarrow t}$.}
\vspace{0mm}
\label{fig:overview}
\end{figure}

\subsection{Cage-based deformations}\label{sec:mvc}
CBD are a type of freeform space deformations. Instead of defining a deformation solely on the surface $ \sh $, space deformations warp the entire ambient space in which the shape $\sh$ is embedded.
In particular, a CBD controls this warping via a coarse triangle mesh, called a \emph{cage} $ \cage $, which typically encloses $ \sh $.
Given the cage, any point in ambient space $\pp\in\mathds{R}^3$ is encoded via generalized barycentric coordinates, as a weighted average of the cage vertices $ \cagev $:  $\pp = \sum \phi^\cage_j\left(\pp\right) \cagev$, where
the weight functions $\left\{\phi^{\cage}_j\right\}$ depend on the relative position of $ \pp $ \wrt to the cage vertices $ \left\{\cagev\right\} $.
The deformation of any point in ambient space  is obtained by simply offsetting the cage vertices and interpolating their new positions $ \cagev' $ with the pre-computed weights, \ie
\vspace{-2mm}
\begin{equation}\label{eq:mvc}
\pp' = \sum_{0\leq j<|V_\cage|}\phi^\cage_j\left(\pp\right)\cagev'.
\vspace{-1mm}
\end{equation}
Previous works on CBD constructed various formulae to attain weight functions $\left\{\phi^\cage_j\right\}$ with specific properties, such as interpolation, linear precision, smoothness and distortion minimization.
We choose mean value coordinates (MVC) \cite{ju2005mean} for their feature preservation and interpolation properties, as well as simplicity and differentiability \wrt the source and deformed cages' coordinates.

\subsection{Learning cage-based deformation}
As our goal is an end-to-end pipeline for deforming shapes, we train the network to predict both the source cage and the target cage, in order to optimize the quality of the resulting deformation.
Given a source shape $ \sh_s $ and a target shape $ \sh_t $, we design a deep neural network that  predicts a cage deformation that warps $ \sh_s $ to $ \sh_t $ while preserving the details of $ \sh_s $.
Our network is composed of two branches, as illustrated in Fig~\ref{fig:overview}:
a cage-prediction model $\net_c$, which predicts the initial cage $ \cage_s $ around $ \sh_s $,
and a deformation-prediction model $ \net_d $, which predicts an offset from $ \cage_s $, yielding the deformed cage $ \cage_{s\to t}$, \ie
\vspace{-1mm}
\begin{equation}
\cage_s = \net_c\left(\sh_s\right) + \cage_0, \qquad \cage_{s\to t} = \net_d\left(\sh_t, \sh_s\right) + \cage_s
\vspace{-1mm}
\end{equation}
Since both branches are differentiable, they can be both learned jointly in an end-to-end manner.

The branches $\net_c$ and $\net_d$ only predict the cage and do not directly rely on the detailed geometric features of the input shapes. Hence, our network does not require high-resolution input nor involved tuning for the network architectures.
In fact, both $ \net_c $ and $ \net_d $ follow a very streamlined design:
their encoders and decoders are simplified versions of the ones used in AtlasNet~\cite{groueix2018atlasnet}. We remove the batch normalization and reduce the channel sizes, and instead of feeding 2D surface patches to the decoders, we feed a template cage $ \cage_0 $ and the predicted initial cage $ \cage_s $ to the the cage predictor and deformer respectively, and let them predict the offsets. By default, $ \cage_0 $ is a 42-vertex sphere.

\subsection{Loss terms}

Our loss  incorporates three main terms. The first term optimizes the source cage to encourage positive mean value coordinates. The two latter terms optimize the deformation, the first by measuring alignment to target and the second by measuring shape preservation. Together, these terms comprise our basic loss function:
\vspace{-2.5mm}
\begin{equation}
\label{eq:main}
\LL = \alpha_\text{MVC}\LL_{\text{MVC}} + \LL_{\text{align}} + \alpha_\text{shape}\LL_{\text{shape}}.
\vspace{-2mm}
\end{equation}
We use $\alpha_\text{MVC}=1$, $\alpha_\text{shape}=0.1$ in all experiments.

To optimize the mean value coordinates of the source cage, we penalize negative weight values, which emerge when the source cage is highly concave, self-overlapping, or when some of the shape's points lie outside the cage:
\vspace{-2mm}
\begin{equation}
\LL_{\text{MVC}} = \frac{1}{|\cage_s||\sh_s|}\sum_{i=1}^{|\sh_s|}\sum_{j=1}^{|\cage_s|}\left|\min\left(\phi_{ji}, 0\right)\right|^2,
\vspace{-2mm}
\end{equation}
where $ \alpha_\text{MVC} $ is the loss weight, and $ \phi_{ji} $ denotes the coordinates of  $ \shapev\in\sh_s $ \wrt  $ \cagev\in\cage_s $.

$\LL_{\text{align}}$ is measured either via chamfer distance in the unsupervised case sans correspondences, or as the L2 distance when supervised with correspondences.
%in an supervised setting if correspondence the source and the target is not known:
%\begin{align}
%\LL_{\text{align}} = & \frac{1}{|\sh_t|}\sum_{\tshapev\in\sh_t}\min_{\shapev'\in\sh_{s\to t}}\|\shapev'-\tshapev\|^2+\\\nonumber
%&\frac{1}{|\sh_{s\to t}|}\sum_{\shapev'\in\sh_{s\to t}}\min_{\tshapev\in\sh_t}\|\shapev'-\tshapev\|^2.
%\end{align}

The above two losses drive the deformation towards alignment with the target, but this may come at the price of preferring alignment over feature preservation. Therefore, we add terms that encourage shape preservation.
Namely, we draw inspiration from Laplacian regularizers~\cite{groueix20183d,wang20193dn,Liu_2019_ICCV}, but  propose to use a point-to-surface distance as an orientation-invariant, second-order geometric feature.
Specifically, for each point $ \pp$ on the source shape, we fit a PCA plane to a local neighborhood $\B$ (we use the one-ring of the mesh), and then compute the point-to-plane distance as $d = \|\nn^T\left(\pp-\pp_\B\right)\|$, where $ \nn $ denotes the normal of the PCA plane and $ \pp_\B=\frac{1}{|\B|}\sum_{\qq\in\B\left(\pp\right)}\qq $ is the centroid of the local neighborhood around $ \pp $. We then penalize change in the distance  $d_i$ for each vertex on the surface:
\vspace{-2mm}
\begin{equation}
\LL_{\text{p2f}} = \frac{1}{|\sh_s|}\sum_{i=1}^{|\sh_s|}\|d_i - d_i'\|^2
\vspace{-2mm}
\end{equation}
where $d_i'$ is the distance post deformation.
In contrast to the uniform Laplacian, which considers the distance to the centroid and hence yields a non-zero value whenever the local neighborhood is not evenly distributed, the proposed point-to-surface distance better describes the local geometric features.
\begin{figure}[t!]
\newcommand{\imgsizesynth}{0.158}
\newcommand{\imgsizesynthsmall}{0.15}
\renewcommand{\arraystretch}{0.1}
\scriptsize\centering
\setlength{\tabcolsep}{0pt}
\begin{tabular}{c|ccccc}
%\begin{tabular}{*{6}{m{0.165\linewidth}}}
& \multicolumn{5}{c}{Target} \\
\makecell{\vspace{1cm}\\Source} &
\makecell{\hspace{0.2cm}\includegraphics[width=\imgsizesynthsmall\linewidth, trim=25ex 45ex 20ex 10ex, clip]{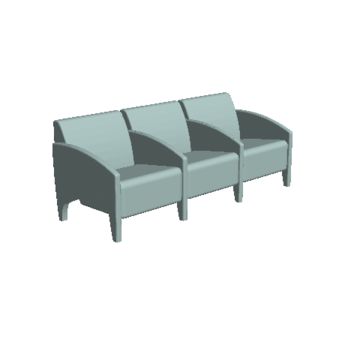}} & %test
\makecell{\includegraphics[width=\imgsizesynthsmall\linewidth]{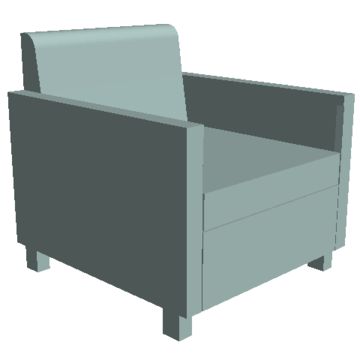}} & %test
\makecell{\includegraphics[width=\imgsizesynthsmall\linewidth, trim=25ex 20ex 20ex 10ex, clip]{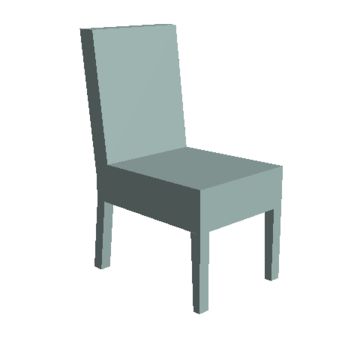}} & %test
\makecell{\includegraphics[width=\imgsizesynthsmall\linewidth, trim=25ex 20ex 20ex 10ex, clip]{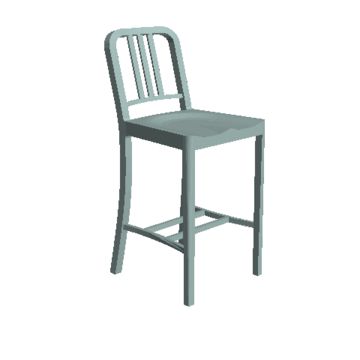}} & %train
\makecell{\includegraphics[width=\imgsizesynthsmall\linewidth, trim=25ex 20ex 20ex 10ex, clip]{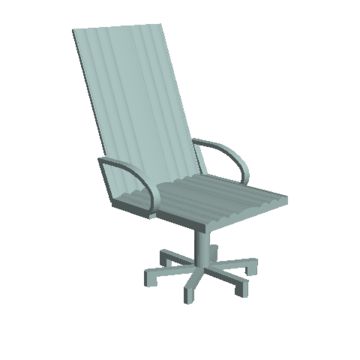}} %test
\\ \midrule
\makecell{\includegraphics[width=\imgsizesynthsmall\linewidth, trim=25ex 25ex 20ex 10ex, clip] {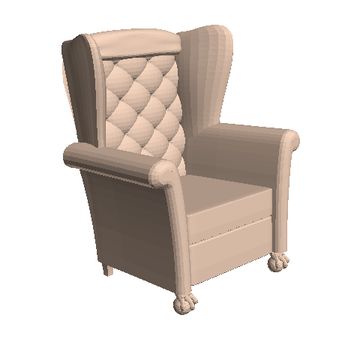}} & %test
\makecell{\includegraphics[width=\imgsizesynth\linewidth, trim=25ex 45ex 30ex 10ex, clip]{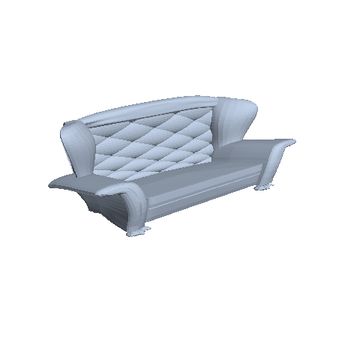}} &
\makecell{\includegraphics[width=\imgsizesynth\linewidth]{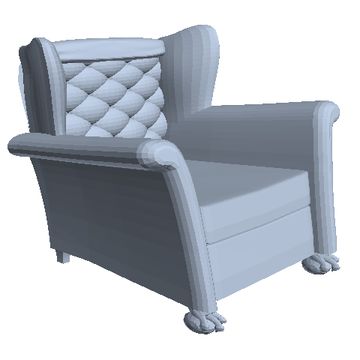}} &
\makecell{\includegraphics[width=\imgsizesynth\linewidth, trim=25ex 25ex 20ex 10ex, clip]{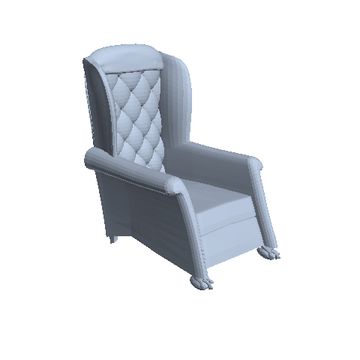}} &
\makecell{\includegraphics[width=\imgsizesynth\linewidth, trim=25ex 25ex 20ex 10ex, clip]{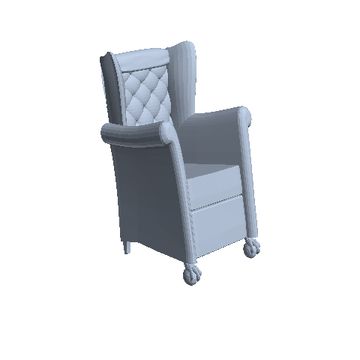}} &
\makecell{\includegraphics[width=\imgsizesynth\linewidth, trim=25ex 15ex 20ex 10ex, clip]{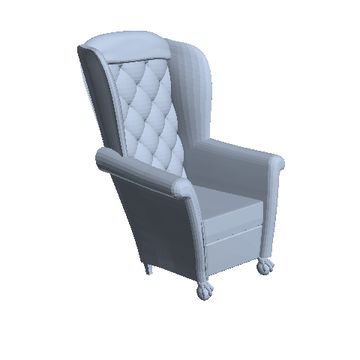}}
\\
\makecell{\includegraphics[width=\imgsizesynthsmall\linewidth, trim=25ex 25ex 20ex 10ex, clip] {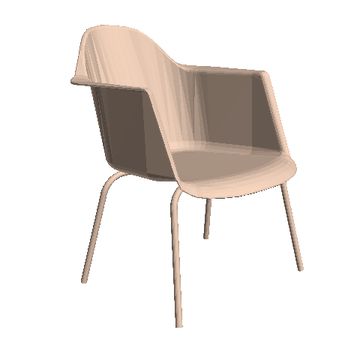}} & %test
\makecell{\includegraphics[width=\imgsizesynth\linewidth, trim=25ex 45ex 30ex 10ex, clip]{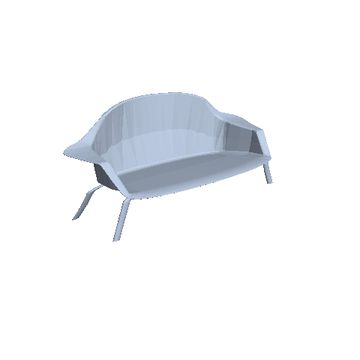}} &
\makecell{\includegraphics[width=\imgsizesynth\linewidth]{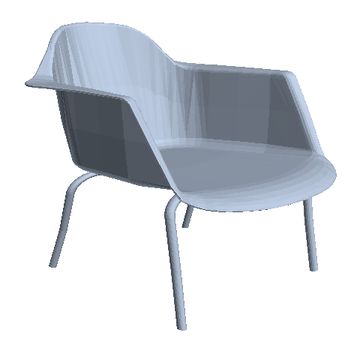}} &
\makecell{\includegraphics[width=\imgsizesynth\linewidth, trim=25ex 25ex 20ex 10ex, clip]{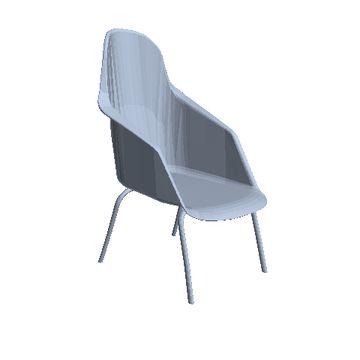}} &
\makecell{\includegraphics[width=\imgsizesynth\linewidth, trim=25ex 25ex 20ex 10ex, clip]{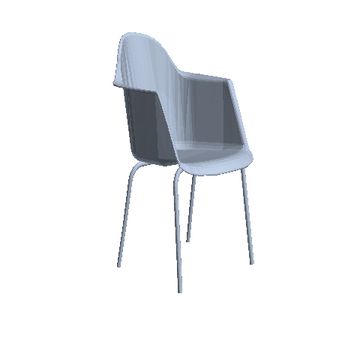}} &
\makecell{\includegraphics[width=\imgsizesynth\linewidth, trim=25ex 25ex 20ex 10ex, clip]{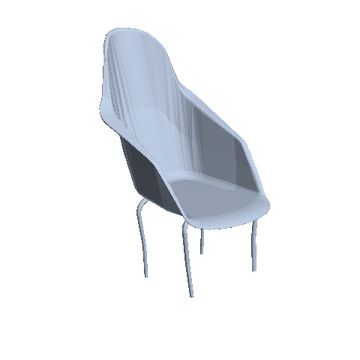}}
\\
\makecell{\includegraphics[width=\imgsizesynthsmall\linewidth, trim=5ex 5ex 10ex 10ex, clip] {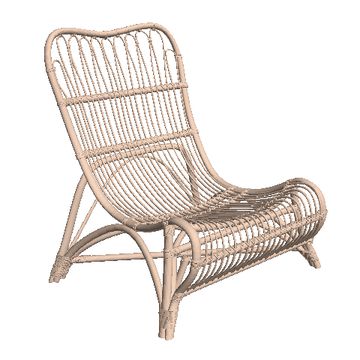}}&
\makecell{\includegraphics[width=\imgsizesynth\linewidth, trim=15ex 25ex 20ex 10ex, clip]{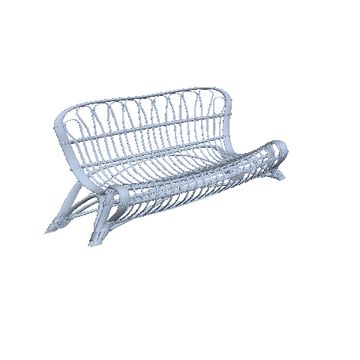}} &
\makecell{\includegraphics[width=\imgsizesynth\linewidth]{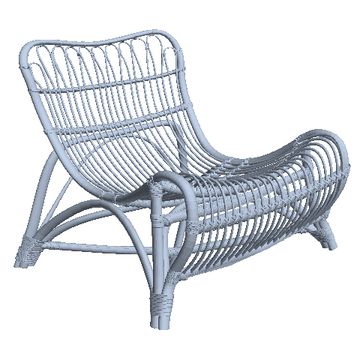}} &
\makecell{\includegraphics[width=\imgsizesynth\linewidth, trim=25ex 25ex 20ex 10ex, clip]{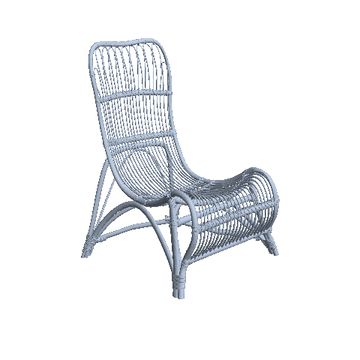}} &
\makecell{\includegraphics[width=\imgsizesynth\linewidth, trim=25ex 25ex 20ex 10ex, clip]{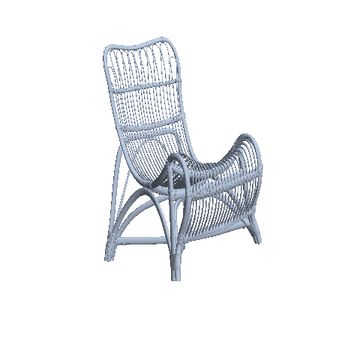}} &
\makecell{\includegraphics[width=\imgsizesynth\linewidth, trim=25ex 25ex 20ex 10ex, clip]{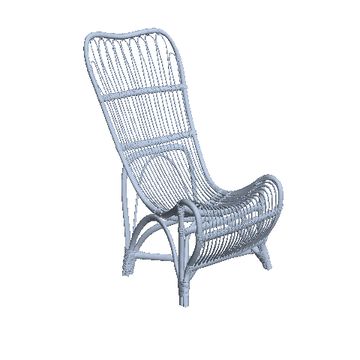}}
\end{tabular}
\vspace{-0.4cm}
\caption{Synthesizing variations of source shapes (brown), by deforming them to match targets (green).}\label{fig:one_to_many}
\vspace{-0.1cm}
\end{figure}

For man-made shapes, we use two additional losses that leverage priors of this shape class.
First, normal consistency is important for, \eg, preserving the planarity of elements like tabletops.
To encourage this, we penalize the angular difference of PCA normals before and after deformation:
%Thus we propose a loss to penalize changes in normals. Specifically, we penalize the angle of the PCA normal before and after the deformation:
\vspace{-1.5mm}
\begin{equation}
\LL_{\text{normal}} = \frac{1}{|\sh_s|}\sum_{i}^{|\sh_s|}(1-\nn_i^T\nn_i'),
\vspace{-2mm}
\end{equation}
where $ \nn' $ denotes the PCA-normal after the deformation.
As demonstrated later, this normal penalty considerably improves the perceptual quality of the deformation.
Second, similarly to Wang \etal\cite{wang20193dn}, we also use the symmetry loss $\LL_{\text{symm}} $, measured as the chamfer distance between the shape and its reflection around the $x=0$ plane. We apply this loss to the deformed shape $ \sh_{s\to t}$ as well as the cage $ \cage_s $. Thus, our final shape preservation loss is: $\LL_\text{shape}\!=\!\LL_\text{p2f}\!+\!\LL_\text{normal}\!+\!\LL_\text{symm}$ for man-made shapes and $\LL_\text{shape}\!=\!\LL_\text{p2f}$ for characters.

%% file: tex/application.tex
%!TEX root = ../main.tex

\begin{figure*}[t!]
\centering\setlength{\tabcolsep}{0pt}\scriptsize
\begin{tabular}{*{7}{p{0.13\linewidth}}}
\makecell{Source} & \makecell{Target} & \makecell{Ours} & \makecell{non-rigid ICP\cite{huang2018learning}} & \makecell{CC\cite{groueix19cycleconsistentdeformation}} & \makecell{3DN\cite{wang20193dn}} & \makecell{ALIGNet\cite{hanocka2018alignet}} \\\hline
\end{tabular}\vspace{0.5ex}
\scriptsize
\includegraphics[width=0.9\linewidth, clip, trim={0cm, 11ex, 0cm, 0cm}] {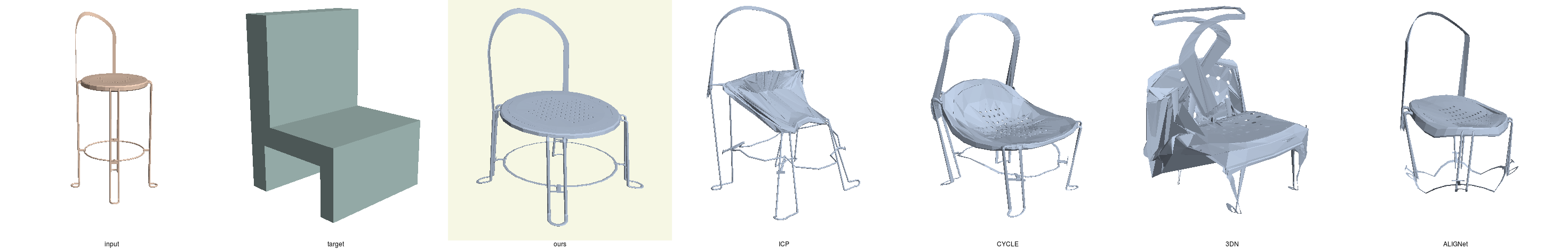}
\includegraphics[width=0.9\linewidth, clip, trim={0cm, 5ex, 0cm, 0cm}]{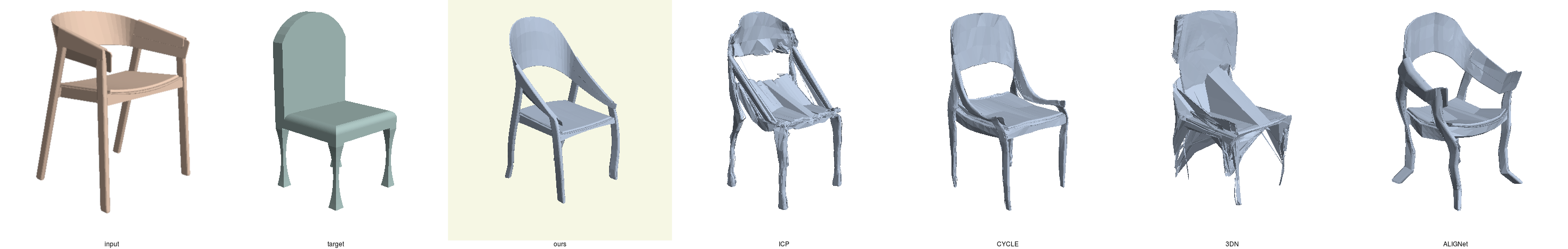}
\includegraphics[width=0.9\linewidth, clip, trim={0cm, 30ex, 0cm, 25ex}]{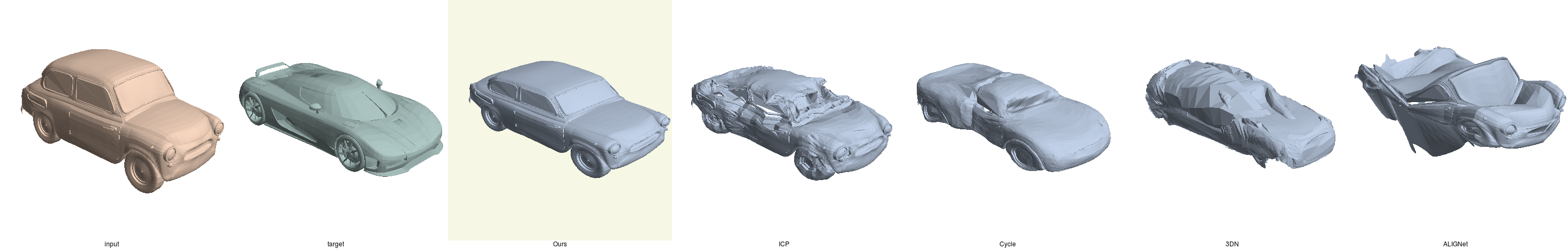}
\includegraphics[width=0.9\linewidth, clip, trim={0cm, 30ex, 0cm, 25ex}]{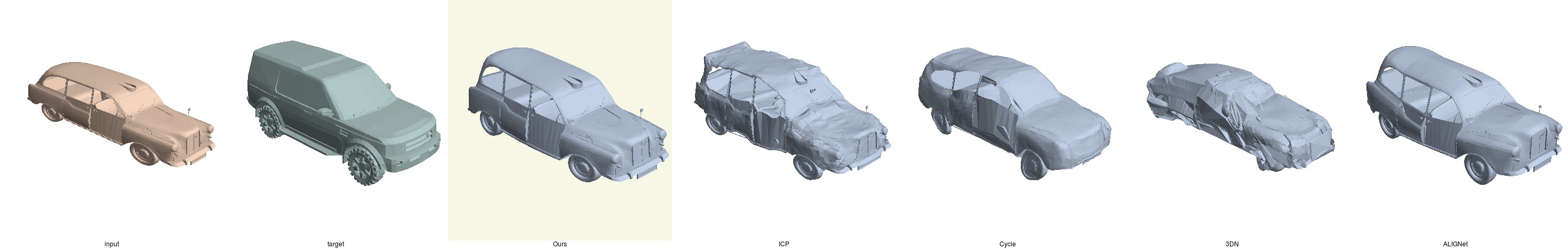}
\includegraphics[width=0.9\linewidth, clip, trim={0cm, 30ex, 0cm, 10ex}] {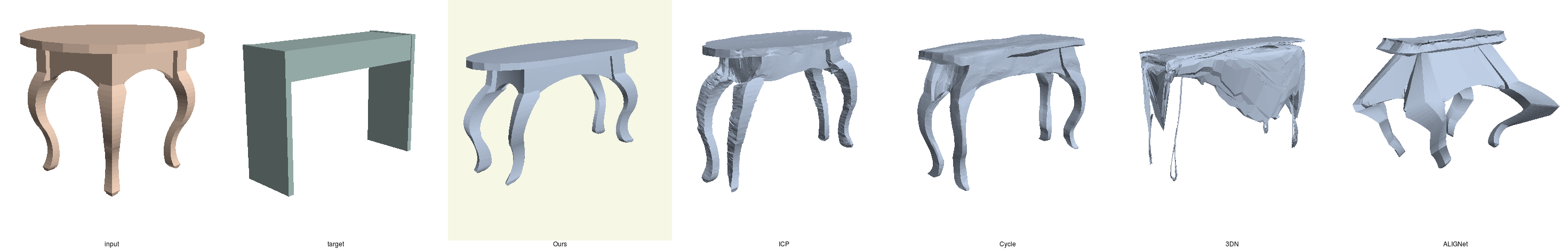}
\includegraphics[width=0.9\linewidth, clip, trim={0cm, 50ex, 0cm, 25ex}] {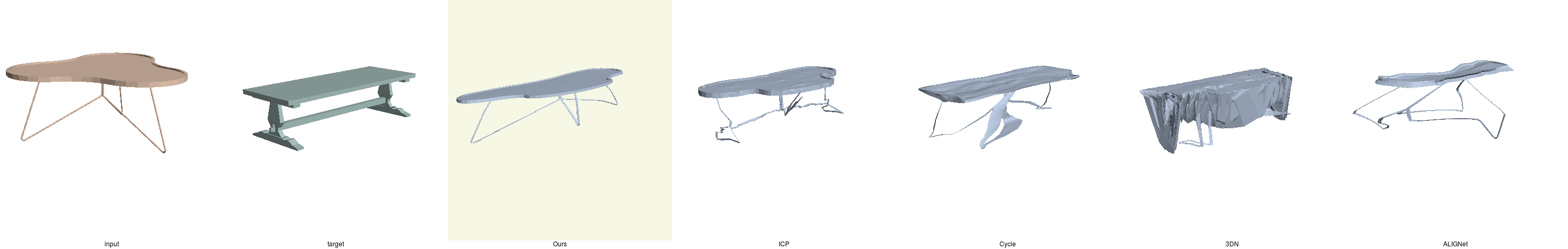}
\vspace{-0.2cm}
\caption{Comparison of our method with other non-homogeneous deformation methods. Our method achieves superior detail preservation of the source shape in comparison to optimization-based
~\cite{huang2018learning} and learning-based~\cite{groueix19cycleconsistentdeformation,wang20193dn,hanocka2018alignet} techniques, while still aligning the output to the target.
}\label{fig:3Dsynthesis}
\vspace{-5mm}
\end{figure*}

\section{Applications}\label{sec:app}
We now showcase two applications of the trained cage-based deformation network.

\input{tex/app_synthesis}

\input{tex/app_deformtransfer}

%% file: tex/app_synthesis.tex
%!TEX root = ../06035.tex
\subsection{Stock amplification via deformation }
Creating high-quality 3D assets requires significant time, technical expertise, and artistic talent. Once the asset is created, the artist commonly deforms the model to create several variations of it. Inspired by prior techniques on automatic stock amplification~\cite{wang20193dn}, we use our method to learn a meaningful deformation space over a collection of shapes within the same category, and then use random pairs of source and target shapes to synthesize plausible variations of artist-generated assets.

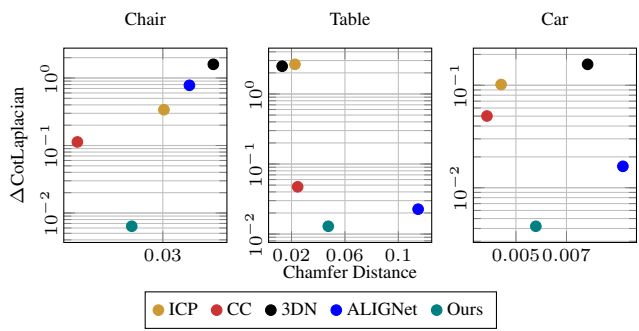
\begin{figure}[t!]
	\scriptsize\hspace{-3ex}
	\begin{tikzpicture}
	\begin{groupplot}[
	group style={group size=3 by 1, horizontal sep=5ex,},
	ymode=log,
	legend style={/tikz/every even column/.append style={column sep=1ex}},
	xticklabel style={scaled ticks=false, /pgf/number format/precision=3, /pgf/number format/fixed},
	%\begin{semilogyaxis}[
	yticklabel style={rotate=90},
	height=0.5\linewidth,
	width=0.45\linewidth,
	grid=both,
	legend pos=outer north east,
	scatter/classes={
		%scaling={purple},
		icp={orange},
		cc={red},
		3dn={black},
		alig={blue},
		ours={teal}
	}]

	\nextgroupplot[ylabel={$ \Delta $CotLaplacian},title=Chair,
	legend to name={CommonLegend},legend style={legend columns=6},
	xtick={0.03, 0.06},
	]
	\addplot[scatter,only marks,
	scatter src=explicit symbolic,
	]
	coordinates {
		%(0.06034, 0.002955) [scaling]
		(0.0302, 0.3389)  [icp]
		(0.01428, 0.1129) [cc]
		(0.03929, 1.595)  [3dn]
		(0.03489, 0.7798) [alig]
		(0.02428, 0.006359) [ours]
	};
	\legend{%Anisotropic Scaling,
		ICP, CC, 3DN, ALIGNet, Ours}
	\nextgroupplot[xlabel={Chamfer Distance},title=Table, xlabel shift=-1ex,
	xtick={0.02, 0.06, 0.1},
	]
	\addplot[scatter,only marks,
	scatter src=explicit symbolic,
	]
	coordinates {
		%(0.07018, 0.003411) [scaling]
		(0.02258, 2.649)  [icp]
		(0.0246, 0.04731) [cc]
		(0.013, 2.489)	[3dn]
		(0.1148, 0.02267) [alig]
		(0.04754, 0.01292) [ours]
	};
	\nextgroupplot[title=Car,
	xtick={0.005, 0.007},
	]
	\addplot[scatter,only marks,
	scatter src=explicit symbolic
	]
	coordinates {
		%(0.007944, 0.003088) [scaling]
		(0.004413, 0.1017) [icp]
		(0.003853, 0.0501) [cc]
		(0.007836, 0.1601) [3dn]
		(0.009232, 0.01618) [alig]
		(0.005788, 0.004204) [ours]
	};
	\end{groupplot}
	\node[yshift=0ex, below] at (current bounding box.south)
	{\pgfplotslegendfromname{CommonLegend}};
	\end{tikzpicture}
	\vspace{-3mm}
	\caption{Quantitative evaluation of our method vs alternative methods. Each point represents a method, embedded according to its average alignment error (Chamfer Distance) and distortion ($\Delta$CotLaplacian). Points near the bottom-left corners are better.}\label{fig:3Deval}
	\vspace{-1mm}
\end{figure}
\noindent \textbf{Training details.} We train our model on the \emph{chair, car} and \emph{table} categories from ShapeNet~\cite{shapenet2015} using the same splitting into training and testing sets as in Grouiex \etal~\cite{groueix19cycleconsistentdeformation}. We then randomly sample 100 pairs  from the test set. Each shape is normalized to fit in a unit bounding box and is represented by 1024 points.

\noindent \textbf{Variation synthesis examples.}
Fig~\ref{fig:one_to_many} shows variations generated from various source-target pairs, exhibiting the regularizing power of the cages: even though our training omits all semantic supervision such as part labels, these variations are plausible and do not exhibit feature distortions; fine details, such as chair slats, are preserved.

\noindent \textbf{Comparisons.}
We compared our target-driven deformation method to other methods that strive to achieve the same goal. Results are shown in Fig~\ref{fig:3Dsynthesis}. While in many cases alternative techniques do align the deformed shape the target, in all cases they introduce significant artifacts in the deformed meshes.

 We first compare to a non-learning-based approach: non-rigid ICP~\cite{huang2018learning}, a classic registration technique that alternates between correspondence estimation and optimization of a non-rigid deformation to best align corresponding points. We show results with the optimal registration parameters we found to achieve detail preservation. Clearly, ICP is sensitive to wrong correspondences that cause convergence to artifact-ridden local minima. We also compare to learning-based methods that directly predict per-point transformations and leverage cycle-consistency (CC)~\cite{groueix19cycleconsistentdeformation} or feature-preserving regularization (3DN)~\cite{wang20193dn} to learn low-distortion shape deformations. Both methods blur and omit features, while also creating artifacts by stretching small parts. We also compare to ALIGNet~\cite{hanocka2018alignet}, a method that predicts a freeform deformation over a voxel grid, yielding a volumetric deformation of the ambient space similarly to our technique. Contrary to our method, the coarse voxel grid cannot capture the fine deformation of the surface needed to avoid large artifacts. Our training setup is identical to CC, and we retrained 3DN and ALIGNet with the same setup using parameters suggested by the authors.

In Fig~\ref{fig:scaling}  we compare our results to the simplest of deformation methods -- anisotropic scaling, achieved by simply %
\begingroup
\setlength{\columnsep}{3mm}%
\begin{wrapfigure}{r}{0.6\linewidth}\vspace{-5mm}
\centering\setlength{\tabcolsep}{0pt}\scriptsize\renewcommand{\arraystretch}{0.1}
\begin{tabular}{*{4}{m{0.24\linewidth}}}
	\makecell{Source} & \makecell{Target} & \makecell{Ours} & \makecell{Anisotropic\\Scaling} \\\hline
\end{tabular}
\includegraphics[width=0.9\linewidth,trim=15ex 15ex 10ex 0, clip]{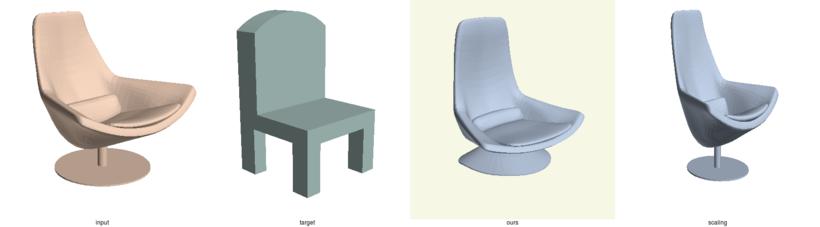}
\includegraphics[width=0.9\linewidth,trim=15ex 15ex 10ex 0, clip]{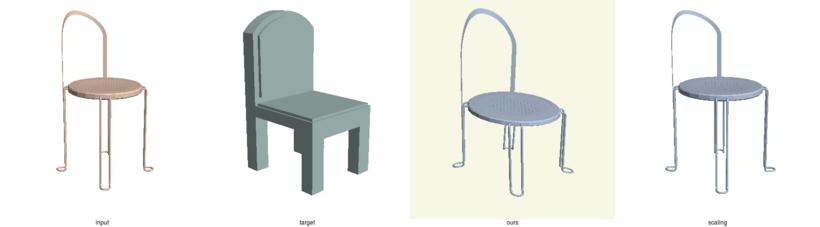}
\includegraphics[width=0.9\linewidth,trim=15ex 15ex 10ex 0, clip]{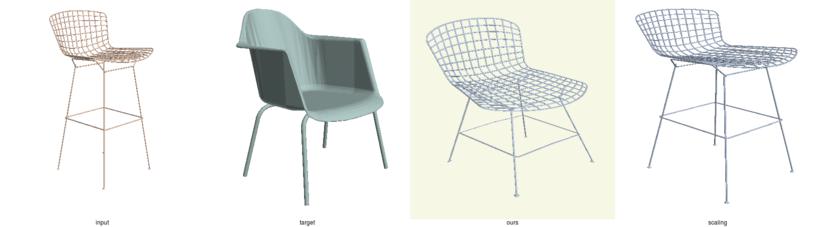}
\vspace{-0.3cm}
\caption{Comparison of our method with anisotropic scaling. Our method better matches corresponding semantic parts.}\label{fig:scaling}
\vspace{-0.2cm}
\end{wrapfigure}
 rescaling the source bounding box to match that of the target. While local structure is well preserved, this method cannot account for the different proportion changes required for different regions, highlighting the necessary intricacy of the optimal deformation in this case.

\endgroup % the preceding blank line is necessary

\noindent \textbf{Quantitative comparisons.}
in Fig~\ref{fig:3Deval}, we quantitatively evaluate the various methods using two metrics: distance to the target shape, and detail preservation, measured via chamfer distance (computed over a dense set of 5000 uniformly sampled points) and difference in cotangent Laplacians, respectively. Note that these metrics do not favor any method, since all optimize for a variant of chamfer distance, and none of the methods optimize for the difference in the cotangent Laplacian. Each 2D point in the figure represents one method, with the point's coordinates prescribed with respect to the two metrics, the origin being ideal. This figure confirms our qualitative observations: our method is more effective at shape preservation than most alternatives while still capturing the gross structure of the target.

\begin{figure}[t!]
\newcommand{\imgsizesvr}{0.165}
\renewcommand{\arraystretch}{0.1}
\centering\setlength{\tabcolsep}{0pt}\small
%\begin{tabular}{*{7}{m{0.14\linewidth}}}
\begin{tabular}{cccc:cc}
\makecell{Target\\Image} & \makecell{Target \\ Proxy~\cite{groueix2018atlasnet}} & \multicolumn{2}{c}{\makecell{Example 1\\Source\hspace{.5cm}Output}} &  \multicolumn{2}{c}{\makecell{Example 2\\Source\hspace{.5cm}Output}}
 \\\midrule
\includegraphics[width=\imgsizesvr\linewidth,clip,trim={0, -0.5cm, 0, 0}]{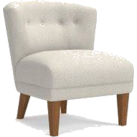}&
\includegraphics[width=\imgsizesvr\linewidth,clip,trim={1cm, 0.5cm, 1cm, 0}]{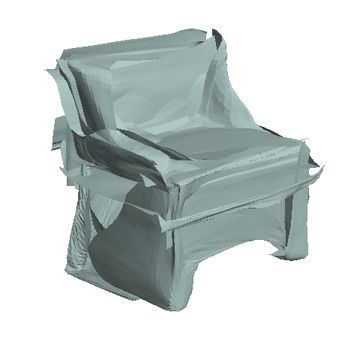}&
\includegraphics[width=\imgsizesvr\linewidth,clip,trim={1cm, 0.5cm, 1cm, 0}]{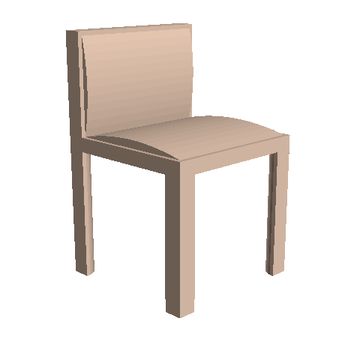}&
\includegraphics[width=\imgsizesvr\linewidth,clip,trim={1cm, 0.5cm, 1cm, 0}]{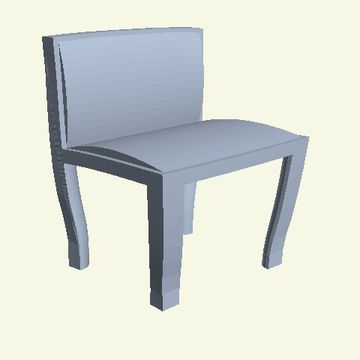}&
\includegraphics[width=\imgsizesvr\linewidth,clip,trim={1cm, 0.5cm, 1cm, 0}]{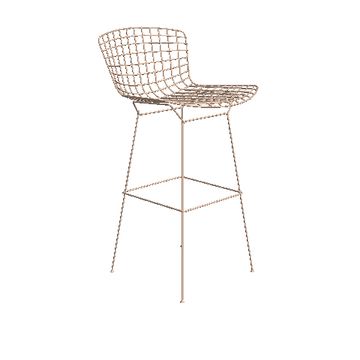}&
\includegraphics[width=\imgsizesvr\linewidth,clip,trim={1cm, 0.5cm, 1cm, 0}]{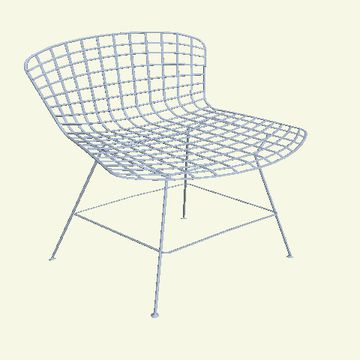}\\
\includegraphics[width=\imgsizesvr\linewidth,clip,trim={0, 0.5cm, 0, 0}]{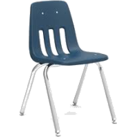}&
\includegraphics[width=\imgsizesvr\linewidth,clip,trim={0, 0.5cm, 0, 0}]{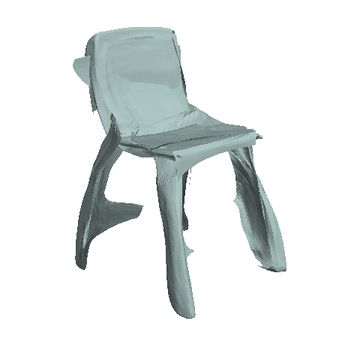}&
\includegraphics[width=\imgsizesvr\linewidth,clip,trim={0, 0.5cm, 0, 0}]{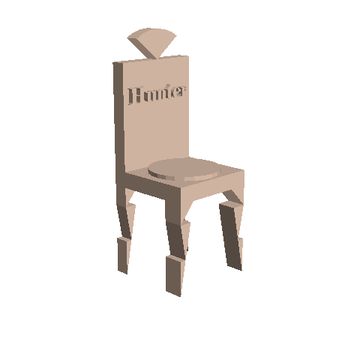}&
\includegraphics[width=\imgsizesvr\linewidth,clip,trim={0, 0.5cm, 0, 0}]{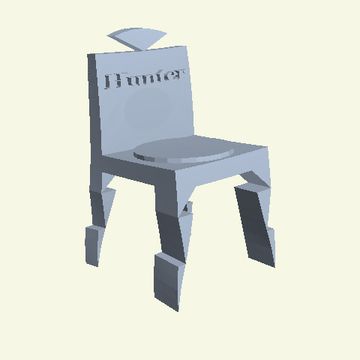}&
\includegraphics[width=\imgsizesvr\linewidth,clip,trim={0, 0.5cm, 0, 0}]{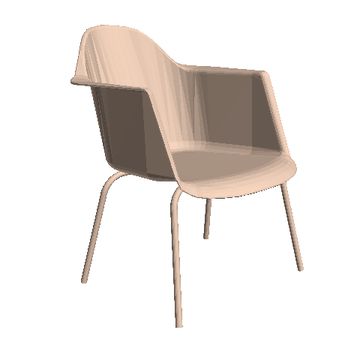}&
\includegraphics[width=\imgsizesvr\linewidth,clip,trim={0, 0.5cm, 0, 0}]{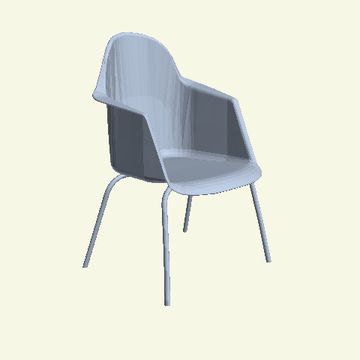}\\
\includegraphics[width=\imgsizesvr\linewidth,clip,trim={0, 0.5cm, 0, 0}]{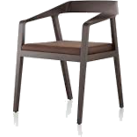}&
\includegraphics[width=\imgsizesvr\linewidth,clip,trim={0, 0.5cm, 0, 0}]{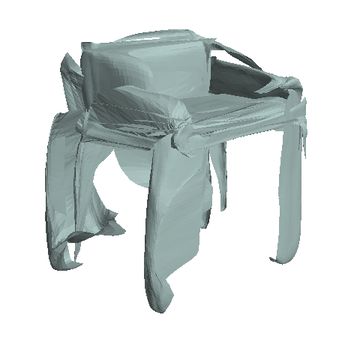}&
\includegraphics[width=\imgsizesvr\linewidth,clip,trim={0, 0.5cm, 0, 0}]{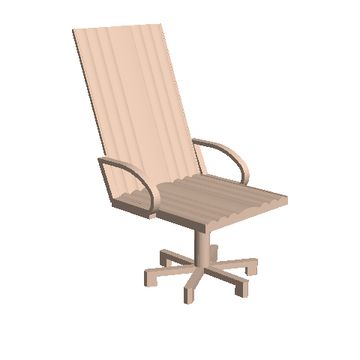}&
\includegraphics[width=\imgsizesvr\linewidth,clip,trim={0, 0.5cm, 0, 0}]{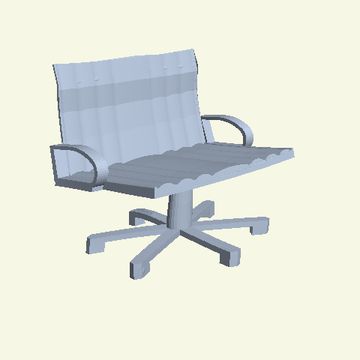}&
\includegraphics[width=\imgsizesvr\linewidth,clip,trim={0, 0.5cm, 0, 0}]{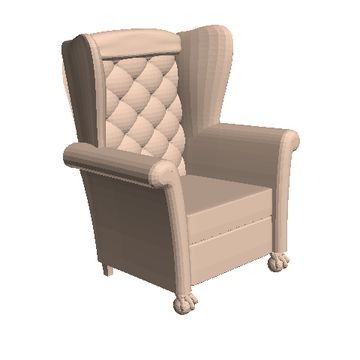}&
\includegraphics[width=\imgsizesvr\linewidth,clip,trim={0, 0.5cm, 0, 0}]{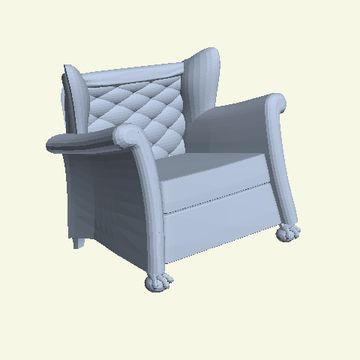}
\end{tabular}
\vspace{-0.2cm}
\caption{We use our method to deform a 3D shape to match a real 2D image. We first use AtlasNet~\cite{groueix2018atlasnet} to reconstruct a 3D proxy target. Despite the poor quality of the proxy, it still serves as a valid target for our network to generate a matching output preserving the fine details of the source. }\label{fig:svr_real}
	\vspace{0mm}
\end{figure}

\noindent \textbf{Using images as targets.}
Often, a 3D target is not readily available. Images are more abundant and much easier to acquire, and thus pose an appealing alternative. We use a learning-based single-view reconstruction technique to create a proxy target to use with our method to find appropriate deformation parameters. We use publicly available product images of real objects and execute AtlasNet's SVR reconstruction~\cite{groueix2018atlasnet} to generate a coarse 3D proxy as a target. Fig~\ref{fig:svr_real} shows that even though the proxy has coarse geometry and many artifacts, these issues do not affect the deformation, and the result is still a valid variation of the source.

%% file: tex/app_deformtransfer.tex
%!TEX root = ../06035.tex

\subsection{Deformation transfer}\label{sec:human}
Given a novel 3D model, it is much more time-efficient to automatically deform it to mimic an existing example deformation, than having an artist deform the novel model directly. This automatic task is called \emph{deformation transfer}. The example deformation is given via a model in a rest pose $\sh_s$, and a model in the deformed pose $\sh_t$. The novel 3D model is given in a corresponding rest post $\sh_{s'}$. The goal is to deform the novel model to a position $\sh_{t'}$ so that the deformation $\sh_{s'}\to\sh_{t'}$ is analogous to $\sh_{s}\to\sh_{t}$. This task can be quite challenging, as the example deformation $\sh_{t}$ may have very different geometry, or even come from an ad-hoc scan, and thus dense correspondences between $\sh_s$ and $\sh_t$ are unavailable, preventing the use of traditional mesh optimization techniques such as~\cite{sumner2004deformtransfer}. Furthermore, as the novel character $\sh_{s'}$ may be significantly different from all models observed during training, it is impossible to a-priori learn a deformation subspace for $\sh_{s'}$ unless sufficient pose variations of $\sh_{s'}$ is available, as in Gao \etal~\cite{gao2018automatic}.

We demonstrate that our learning-based approach can be used to perform deformation transfer on arbitrary humanoid models. The network infers the deformation from the source $\sh_s$ to the target $\sh_t$, without any given correspondences, and then an optimization-based method transfers this deformation to a novel shape $\sh_{s'}$ to obtain the desired deformation $\sh_{t'}$. Hence, given \emph{any} arbitrarily-complex novel character, all our method requires are sparse correspondences supplying the necessary alignment between the two rest poses, $\sh_s$ and  $\sh_{s'}$. We now overview the details of our learned cage-based human deformation model and the optimization technique used to transfer the deformations.

\begin{figure}[t!]
\setlength{\tabcolsep}{0pt}
\centering\scriptsize
\begin{tabular}{*{7}{m{0.14\linewidth}}}
\makecell{Source} & \makecell{Target 1} & \makecell{Deformed 1} & \makecell{Target 2} & \makecell{Deformed 2}&\makecell{Target 3}&\makecell{Deformed 3}\\%\midrule
\vspace{-.3cm}
\makecell{\includegraphics[width=\linewidth, trim=0 0 0 2ex, clip]{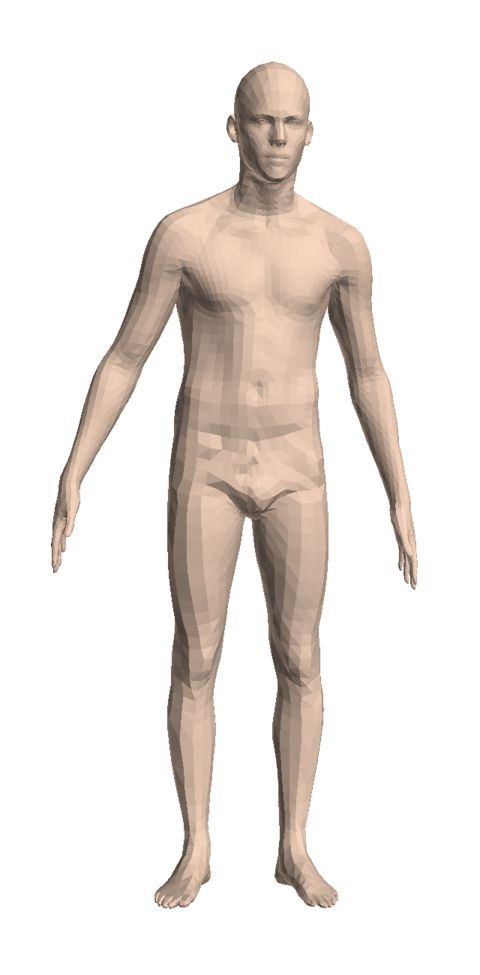}}&
\vspace{-.3cm}
\makecell{\includegraphics[width=\linewidth]{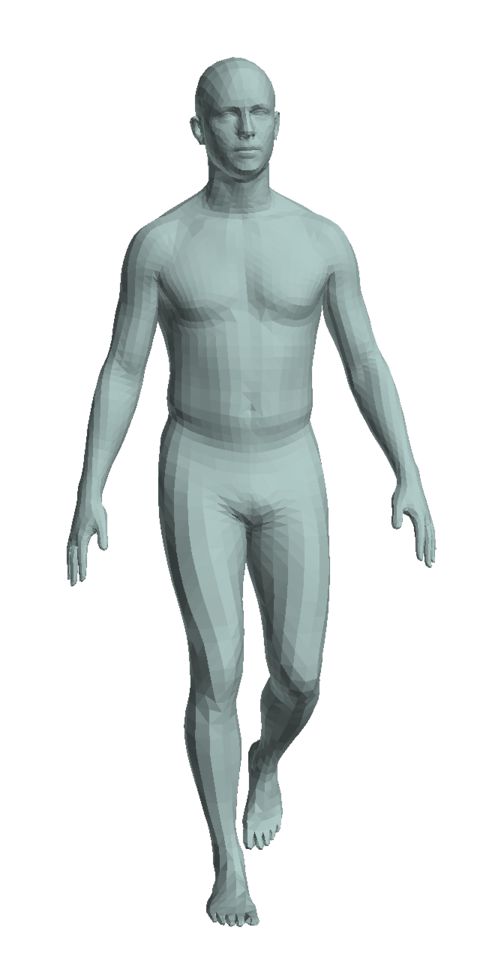}}&
\vspace{-.3cm}
\makecell{\includegraphics[width=\linewidth]{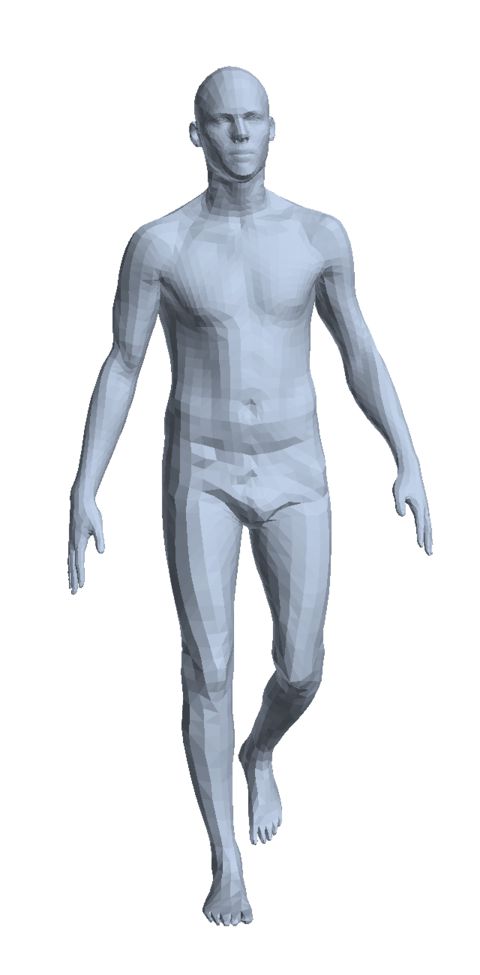}}&
\vspace{-.3cm}
\makecell{\includegraphics[width=\linewidth]{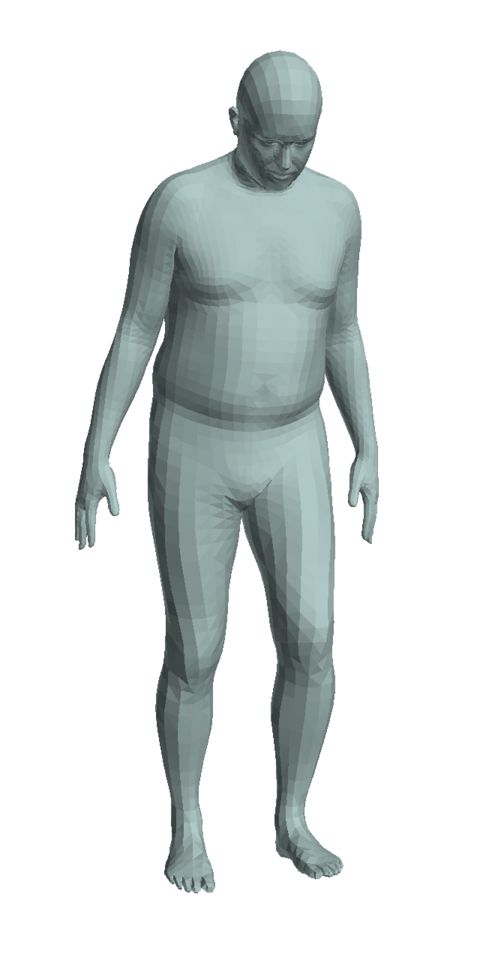}}&
\vspace{-.3cm}
\makecell{\includegraphics[width=\linewidth]{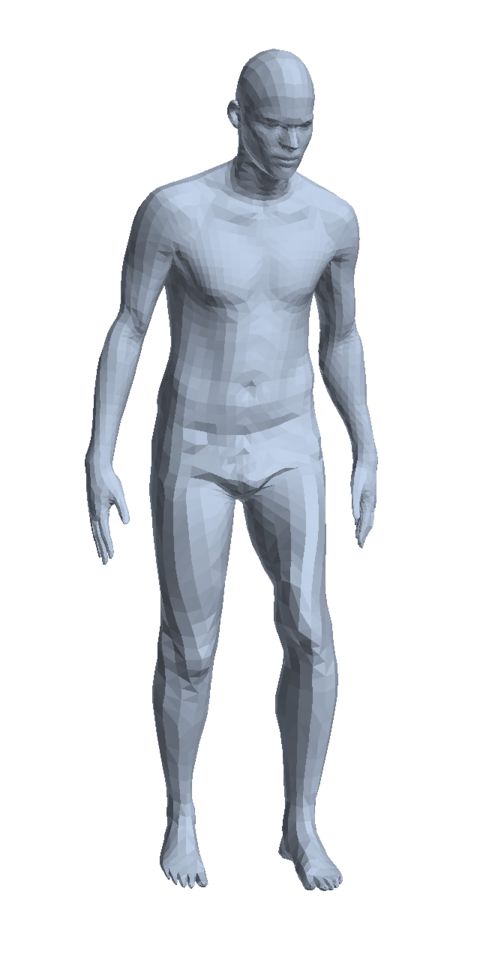}}&
\vspace{-.3cm}
\makecell{\includegraphics[width=\linewidth]{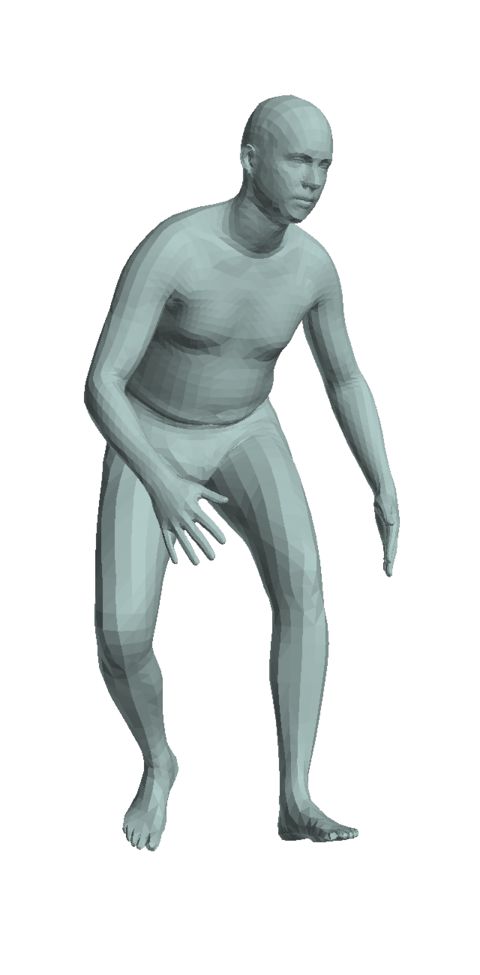}}&
\vspace{-.3cm}
\makecell{\includegraphics[width=\linewidth]{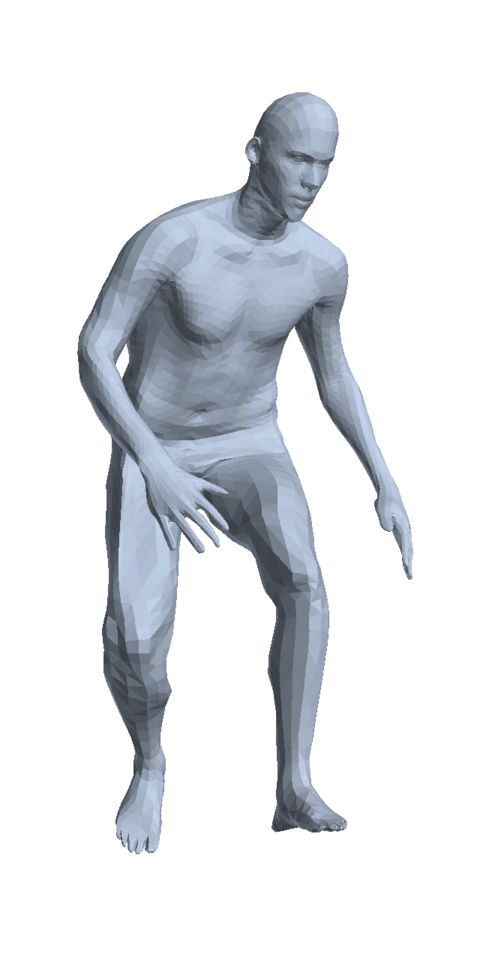}}
\end{tabular}
\vspace{-5mm}
\caption{The deformation model, trained to deform a fixed source (left) to various articulations.}\label{fig:dt_trained}
\vspace{-3mm}
\end{figure}

\noindent \textbf{Learning cage-based human deformation.}
To train our human-specific deformation model, we use the dataset ~\cite{groueix20183d} generated using the SMPL model~\cite{bogo2014faust} of 230K models of various humans in various poses. Since our application assumes that the exemplar deformation is produced from a single canonical character, we picked one human in the dataset to serve as $\sh_s$. 
Subsequently, since we only have one static source shape $\sh_s$, we  use a \emph{static} cage $\cage_s$ manually created with 77 vertices, and hence do not need the cage prediction network $\net_c$ and only use the deformation network  $\net_d$. We train $\net_d$ to deform the static $\sh_s$ using the static $\cage_s$ into exemplars $\sh_t$ from the dataset (with targets not necessarily stemming from the same humanoid model as $\sh_s$). We then train with the loss in \eqref{eq:main}, but with one modification: in similar fashion to prior work, during training we use ground truth correspondences and hence replace the chamfer distance with the L2 distance w.r.t the known correspondences.
Note that these correspondences are \emph{not} used at inference time.

Lastly, during training we also optimize the static source cage $\cage_c$ by treating its vertices as degrees of freedom and directly optimizing them to reduce the loss so as to attain a more optimal, but still static cage after training.

Fig~\ref{fig:dt_trained} shows examples of human-specific cage deformations predicted for test targets (not observed while training). Note how our model successfully matches poses even without knowing correspondences at inference time, while preserving fine geometric details such as faces and fingers.

\begin{figure}[t!]
\centering
\includegraphics[width=\linewidth]{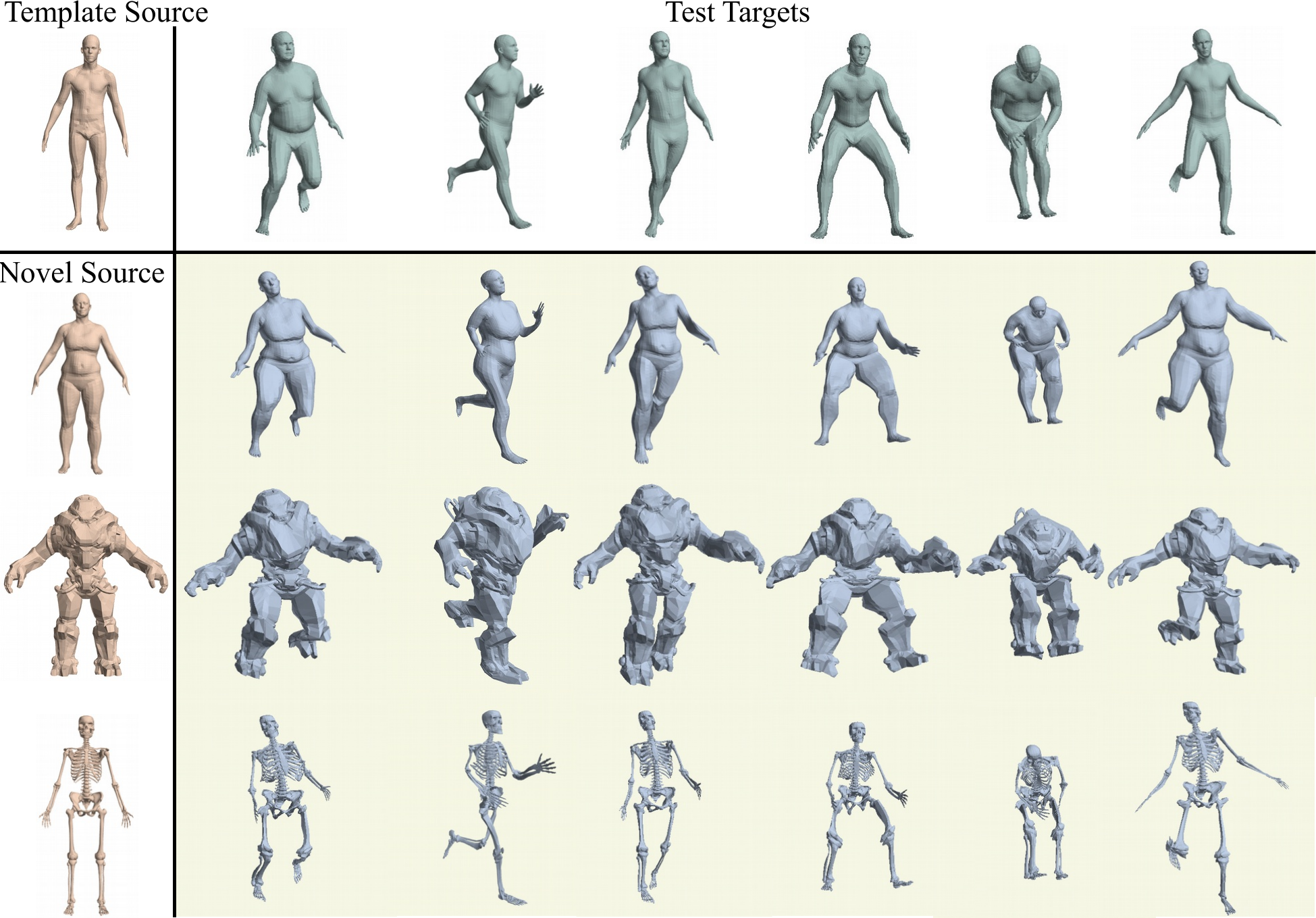}
\vspace{-.5cm}
\caption{Deformation transfer. We first learn the cage deformation space for a template source shape (top left) with known pose and body shape variations. Then, we annotate predefined landmarks on new characters in neutral poses (left column, rows 2-4). At test time, given novel target poses (top row, green) without known correspondences to the template, we transfer their poses to the other characters (blue). 
}
\vspace{0.1cm}
\label{fig:dt_composed}
\end{figure}

\noindent \textbf{Transferring cage deformations.}
After training, we have at our disposal the deformation network $\net_d$ and the static $\cage_s, \sh_s$. We assume to be given a novel character $\sh_{s'}$ with 83 landmark correspondences aligning it to $\sh_s$, and an example target pose $\sh_t$. Our goal is to deform $\sh_{s'}$ into a new pose $\sh_{t'}$ that is analogous to the deformation of $\sh_s$ into $\sh_{t}$.

We first generate a new cage $\cage_{s'}$ for the character $\sh_{s'}$. Instead of a network-based prediction, we simply optimize the static cage $\cage_s$, trying to match mean value coordinates between corresponding points of $\sh_s,\sh_{s'}$:
\vspace{-2mm}
\begin{equation}
\LL_{\text{consistency}} =  \sum_j \sum_{(p,q)} \| \phi_j^{\cage_s}(p)- \phi_j^{\cage_s'}(q)\|^2
\vspace{-2mm}
\end{equation}
where $(p,q)$ are corresponding landmarks. We also regularize with respect to the cotangent Laplacian of the cage:
\vspace{-2mm}
\begin{equation}
\LL_{\cage\text{lap}} =  \sum_{0\leq j<|\cage_s|}\left(\|L_{\text{cot}}\cagev\|-\|L_{\text{cot}}\cagev'\|\right)^2.
\vspace{-1.5mm}
\end{equation}
Then, we compute $\cage_{s'}$ by minimizing $\LL=\LL_{\text{consistency}}+0.05\LL_{\cage\text{lap}}$, with $\cage_s$ used as initialization, solved via the Adam optimizer with step size $5\cdot10^{-4}$ and up to $10^4$ iterations (or until $\LL_{\text{consistency}}<10^{-5}$).

Finally, given the cage $\cage_{s'}$ for the novel character, we compute the deformed cage $\cage_{s'\to t'}$, using our trained deformation network, by applying the predicted offset to the optimized cage: $\cage_{s' \to t'} = \net_d\left(\sh_t, \sh_{s'}\right) +\cage_{s'}$. The final deformed shape $\sh_{t'}$ is computed by deforming $\sh_{s'}$ using the cage $\cage_{s' \to t'}$ via \eqref{eq:mvc}. This procedure is illustrated in Fig~\ref{fig:closeup}, while more examples can be found in the supplemental material.
Due to the agnostic nature of cage-deformations to the underlying shape, we are able to seamlessly combine machine learning and traditional geometry processing to generalize to never-observed characters. To demonstrate the expressiveness of our method, we show examples on extremely dissimilar target characters in Figures~\ref{fig:teaser} and ~\ref{fig:dt_composed}.
\begin{figure}\centering
\includegraphics[width=\linewidth, trim=0 0ex 0 2ex]{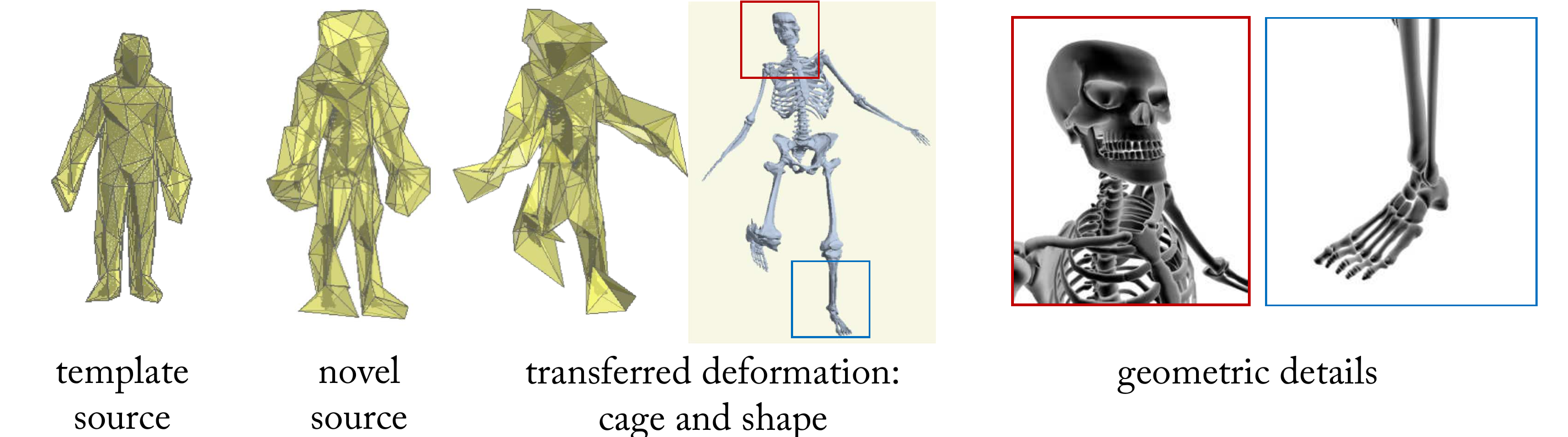}
\vspace{-3.5ex}
\caption{\changed{In deformation transfer, the manually created cage for a template shape (leftmost) is fitted to a novel source shape (second left) by optimizing MVC of a sparse set of aligned landmarks. The learnt deformation can be directly applied to the fitted source cage (columns 3-4), preserving rich geometric features (right).}}\label{fig:closeup}
\end{figure}

%% file: tex/evaluation.tex
%!TEX root = ../06035.tex

\section{Evaluation}
In this section, we study the effects and necessity of the most relevant components of our methods.
To measure the matching error we use chamfer distance computed on 5000 uniformly resampled points, and to measure the feature distortion we use the distance between cotangent Laplacians. All models are normalized to a unit bounding box.

\begin{figure}\scriptsize\centering	\setlength{\tabcolsep}{2pt}\renewcommand{\arraystretch}{0.1}
\begin{tabular}{*{4}{m{0.1\linewidth}}:*{4}{m{0.12\linewidth}}}
\multicolumn{4}{c}{Example 1} & \multicolumn{4}{:c}{Example 2} \\ \toprule
\includegraphics[width=\linewidth,trim=3cm 5ex 4.2cm 0, clip] {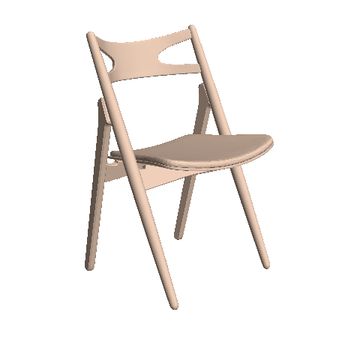}&
\includegraphics[width=\linewidth,trim=3cm 5ex 4.2cm 0, clip] {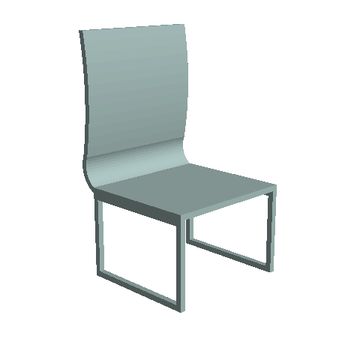}&
\includegraphics[width=\linewidth,trim=3cm 5ex 4.2cm 0, clip] {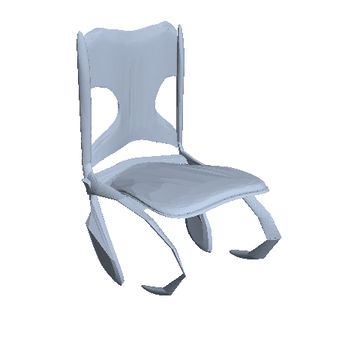}&
\includegraphics[width=\linewidth,trim=3cm 5ex 4.2cm 0, clip] {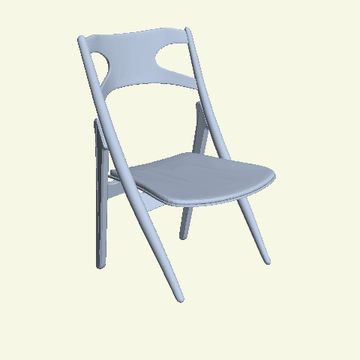} & 
\includegraphics[width=\linewidth,trim=3cm 5ex 3cm 0, clip]{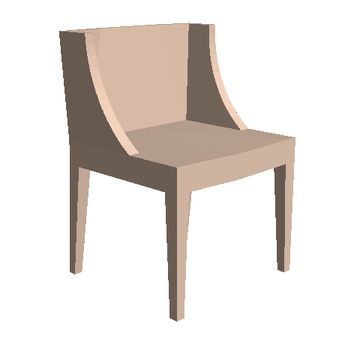}&
\includegraphics[width=\linewidth,trim=3cm 5ex 3cm 0, clip]{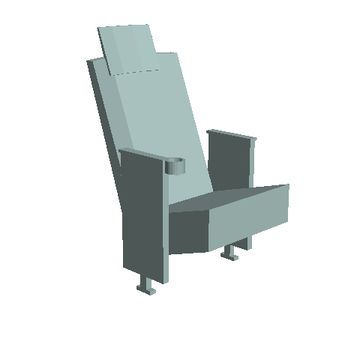}&
\includegraphics[width=\linewidth,trim=3cm 5ex 3cm 0, clip]{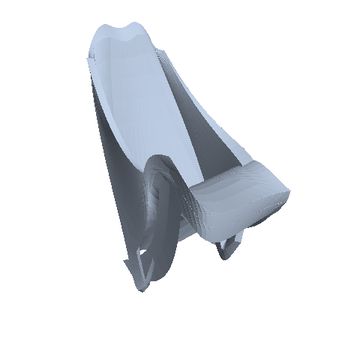}&
\includegraphics[width=\linewidth,trim=3cm 5ex 3cm 0, clip]{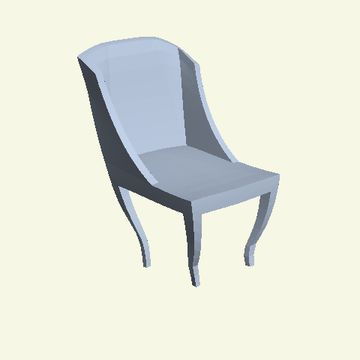}\\
\makecell{Source} & \makecell{Target} & \makecell{Opt.} & \makecell{Ours} & \makecell{Source} & \makecell{Target} & \makecell{Opt.} & \makecell{Ours}
\end{tabular}
\caption{Our approach produces more plausible inter-shape correspondences and deformations than per-pair optimization.}\label{fig:overfit}
\end{figure}

\changed{\noindent \textbf{Benefit of learning CBD from data.}
Instead of learning the CBD from a collection of data, one could minimize \eqref{eq:main} for a single pair of shapes, which is essentially a non-rigid Iterative-Closest-Point (ICP) parameterised by cage vertices.
As shown in Fig~\ref{fig:overfit}, when correct correspondence estimation becomes challenging, the optimization alternative produces non-plausible outputs.
In contrast, the learnt approach utilizes domain knowledge embedded in the network's parameters  \cite{zeiler2014visualizing,shaham2019singan}, amounting to better reasoning about the plausibility of inter-shape correspondences and deformations.
The learned domain knowledge can generalize to new data. As demonstrated
in Sec~\ref{sec:human}, even though our network is trained with ground-truth correspondences, it is able to automatically associate the source shape to a new target \emph{without correspondences} during inference, while optimization methods require accurate correspondence estimation for every new target.}

\begin{figure}[t!]
\includegraphics[width=\linewidth]{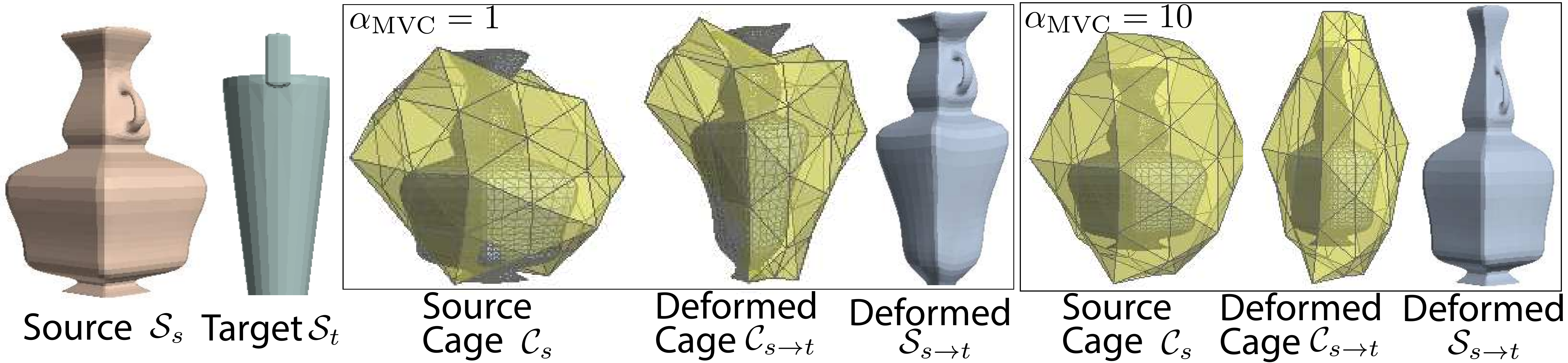}
\vspace{-0.6cm}
\caption{Effect of $\LL_\text{MVC}$. Higher regularization yields more conservative deformations.}\label{fig:ablation-mvc}
\vspace{-0.2cm}
\end{figure}

% MVC regularization
\noindent \textbf{Effect of the negative MVC penalty, $\LL_\text{MVC}$.}
in Fig~\ref{fig:ablation-mvc} we show the effect of penalizing negative mean value coordinates. We train our architecture on 300 vase shapes from COSEG~\cite{wu2014interactive}, while varying the weight $\alpha_\text{MVC} \in \{0, 1, 10\}$. Increasing this term brings the cages closer to the shapes' convex hulls, leading to more conservative deformations. Quantitative results in Table~\ref{tab:ablation}a also suggest that increasing the weight $\alpha_\text{MVC}$ favors shape preservation over alignment accuracy. Completely eliminating this term hurts convergence, and increases the alignment error further.

\input{tex/ablation_tables}

\begin{figure}[t!]\footnotesize
\includegraphics[width=\linewidth,clip,trim={0, 0, 0, 0}] {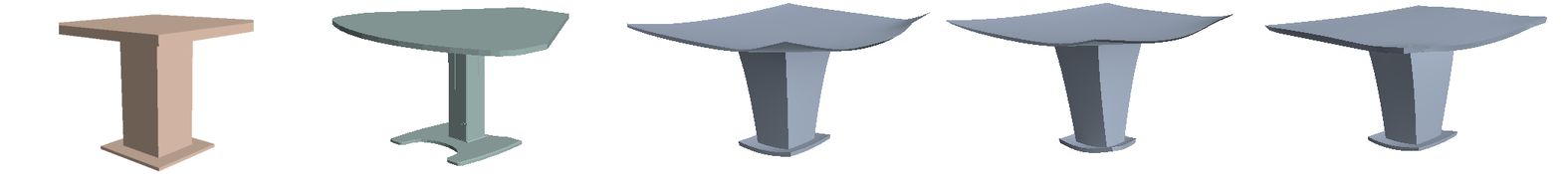}
\includegraphics[width=\linewidth,clip,trim={0, 3cm, 0, 3cm}] {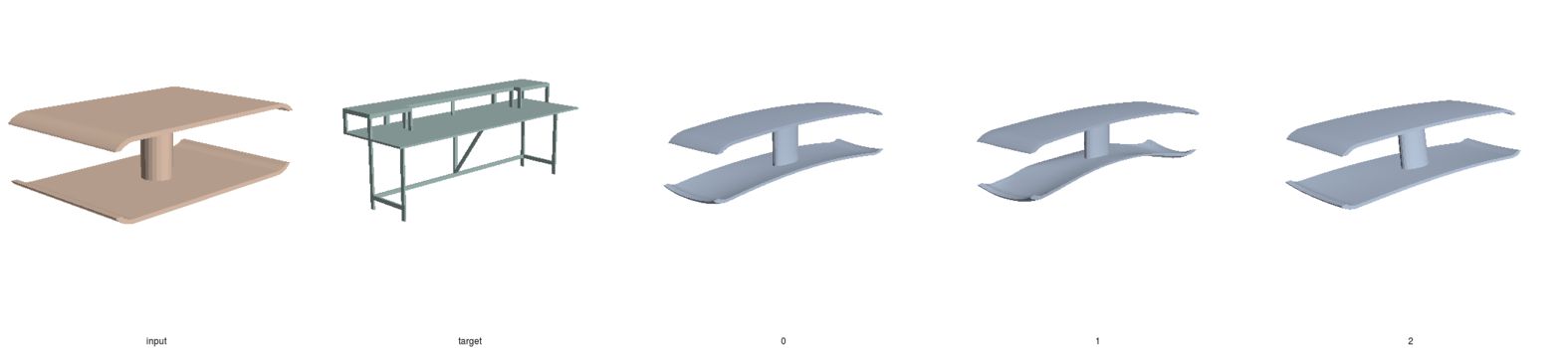}
\begin{tabular}{*{5}{m{0.155\linewidth}}}
\small\setlength{\tabcolsep}{0pt}
\vspace{-7mm}\makecell{Source $ \sh_s $} & \vspace{-7mm}\makecell{Target $ \sh_t $} & \vspace{-7mm}\makecell{$\LL_{\text{lap}} $} & \vspace{-7mm}\makecell{$ \LL_{\text{p2f}} $} & \vspace{-7mm}\makecell{$ \LL_{\text{normal}} $}
\end{tabular}
\vspace{-.9cm}
\caption{The effect of different shape preservation losses, note that all results  include $\LL_\text{symm}$.}\label{fig:ablation-feat}
\vspace{-.2cm}
\end{figure}

\noindent \textbf{Effect of the shape preservation losses, $\LL_\text{shape}$.}
In Fig~\ref{fig:ablation-feat} we compare deformations produced with the full loss ($\LL_\text{shape}=\LL_\text{p2f}+\LL_\text{normal}+\LL_\text{symm}$) to ones produced with only one of the first two loss terms. While we did not use the Laplacian regularizer $\LL_\text{lap}$ as in~\cite{wang20193dn}, it seems to have an effect equivalent to $\LL_\text{p2f}$. As expected, $\LL_{\text{normal}}$ prevents bending of rigid shapes. We quantitatively evaluate these regularizers in Table~\ref{tab:ablation}b, which suggests that $\LL_\text{p2f}$ is slightly better as the deformed shape is more aligned with the target than $\LL_\text{lap}$, even though shape preservation has not been sacrificed.  $\LL_{\text{normal}}$ reduces distortion even further.

\begin{figure}[t!]
\includegraphics[width=\linewidth]{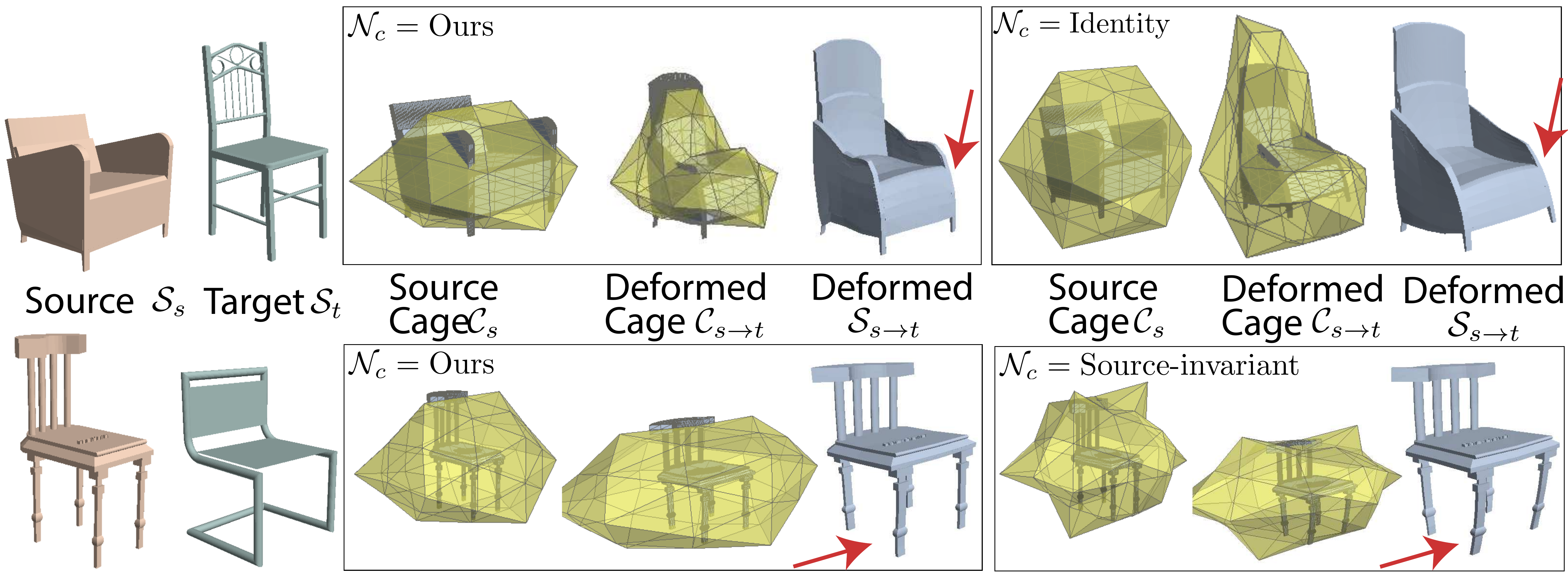}
\vspace{-0.7cm}
\caption{The effect of source-cage prediction. We compare our per-instance prediction of $ \net_c $ with (1) a static spherical cage (top right) and (2) a single optimized cage prediction over the entire training set (bottom right). Our approach achieves better alignment with the target shape.}\label{fig:ablation-cage}
\end{figure}

\noindent \textbf{Design choices for the cage prediction network, $\net_c$.}
The cage prediction network $\net_c$ morphs the template cage mesh (a 42-vertex sphere) into the initial cage enveloping the source shape.  In Fig~\ref{fig:ablation-cage} and Table~\ref{tab:ablation}c we compare to two alternative design choices for this module: an \emph{Identity} module retains the template cage, and a \emph{source-invariant} module in which we optimize the template cage's vertex coordinates with respect to all targets in the dataset, but then use the same fixed cage for testing. Learning source-specific cages produces deformations closest to the target with minimum detail sacrifice. As expected, fixing the template cage produces more rigid deformations, yielding lower distortion at the price of less-aligned results.

%% file: tex/ablation_tables.tex
%!TEX root = ../main.tex

\begin{table}[t!]
\small\centering
\begin{tabular}{lcc}
\toprule
Ablation & CD & $\Delta$CotLaplacian \\ \midrule
$\alpha_\text{MVC}=0$ & 1.64 & 9.04 \\
$\alpha_\text{MVC}=1$ & 1.44 & 8.74 \\
$\alpha_\text{MVC}=10$ & 2.65 & 8.27\\
\multicolumn{3}{c}{(a) Effect of the MVC loss, $\LL_\text{MVC}$} \\ \midrule
$ \LL_\text{shape}\!=\!\LL_{\text{lap}}\!+\!\LL_{\text{symm}} $ & 5.16 & 4.75 \\
$ \LL_\text{shape}\!=\!\LL_{\text{p2f}}\!+\!\LL_{\text{symm}} $ & 4.86 & 4.70\\
$ \LL_\text{shape}\!=\!\LL_{\text{normal}}\!+\!\LL_{\text{symm}} $ & 5.45 & 4.33\\
\multicolumn{3}{c}{(b) Effect of the shape preservation losses, $\LL_\text{shape}$} \\ \midrule
$\net_c$=Identity &  3.27  & 5.65 \\
$\net_c$=Source-invariant & 3.11  & 12.05 \\
$\net_c$=Ours & 3.06 & 10.45 \\
\multicolumn{3}{c}{(c) Design choices for cage prediction network, $\net_c$} \\ \bottomrule
\end{tabular}
\vspace{-0.2cm}
\caption{We evaluate effect of different losses ($\LL_\text{MVC},\LL_\text{shape}$) and components ($\net_c$) of our pipeline with respect to chamfer distance (CD, scaled by $10^2$) and cotangent Laplacian (scaled by $10^3$). }\label{tab:ablation}
\end{table}

%% file: tex/conclusion.tex
%!TEX root = ../main.tex
\section{Conclusion}
We show that classical cage-based deformation provides a low-dimensional, detail-preserving deformation space directly usable in a deep-learning setting. We implement cage weight computation and cage-based deformation as differentiable network layers, which could be used in other architectures. Our method succeeds in generating feature-preserving deformations for synthesizing shape variations and deformation transfer, and better preserves salient geometric features than competing methods.

A limitation of our approach is that we focus on the deformation quality produced by the predicted cages: hence, the cage geometry itself is not designed to be comparable to professionally-created cages for 3D artists. Second, our losses are not quite sufficient to always ensure rectilinear/planar/parallel structures in man-made shapes are perfectly preserved (Fig~\ref{fig:ablation-feat}). \changed{Third, for certain types of deformations other parameterizations might be a more natural choice, such as skeleton-based deformation for articulations, nonetheless the idea presented in this paper can be adopted for similarly. }

Our method provides an extensible and versatile framework for data-driven generation of high detail 3D geometry. In the future we would like to incorporate alternative cage weight computation layers, such as Green Coordinates~\cite{lipman2008green}. Unlike MVC, this technique is not affine-invariant, and thus would introduce less affine distortion for large articulations (see the second row fourth column in Fig~\ref{fig:dt_composed}). We also plan to use our method in other applications such as registration, part assembly, and generating animations.

%% file: 06035-supp.tex
\onecolumn
\section{Supplemental}
In the paper, we detailed two applications of our method: stock amplification and deformation transfer.
In this document, we provide additional results for each application.
\paragraph{Stock amplification via deformation.}\label{sec:shapenet}
In Table~\ref{tab:chairs}, \ref{tab:tables} and \ref{tab:cars} we show the entire deformation results from the \textit{chair}, \textit{table} and \textit{car} categories.
As mentioned in the main paper, the test sets consist of 100 pairs of unseen source and target shapes randomly sampled from ShapeNet dataset~\cite{shapenet2015}.
\begin{figure*}[h!]\centering
	\includegraphics[width=0.8\linewidth]{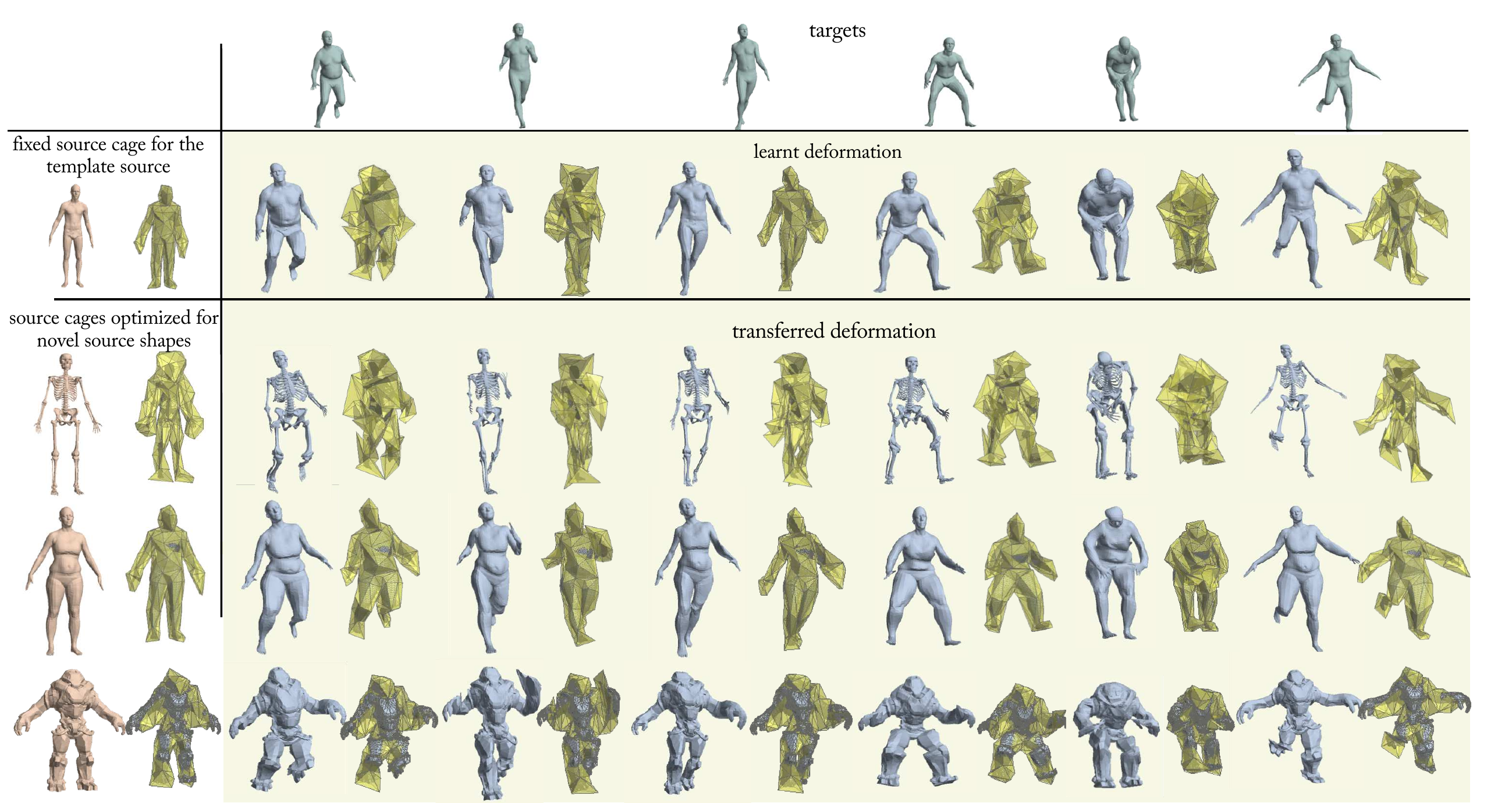}
	\caption{Visualization of the cages used in deformation transfer. Given a template source shape in a known pose with a manually created template cage (second row, left, brown), our deformation network translates the vertices of the template cage to match the source to a novel target shape in various poses (second row, blue). This cage-deformation can be transferred to a new character, provided a pose similar to that of the template source (row 3-5, left, brown).  The deformed novel characters are shown in blue in row 3-5.}\label{fig:cages}
\end{figure*}
\paragraph{Deformation transfer.}\label{sec:dt}
In Figure~\ref{fig:cages} we illustrate the cages used in the humanoid deformation examples.
In Table~\ref{tab:dt} we provide additional results for 100 unseen target poses randomly sampled from the dataset provided by \cite{groueix20183d}.
For each target pose, we show the deformed shape for the original source shape used during training as well as the transferred deformation for three novel source shapes, a woman from FAUST~\cite{bogo2014faust}, a skeleton and a robot.

Two training variations are shown in Table~\ref{tab:dt}. 
On the left, a rest-pose human is used as the source shape during training, whereas on the right a T-pose human is used.
The first variation shows less distortion, but tends to underperform when large arm articulation is present in the target pose.
The second variation shows more distortion on the arms but tends to match the articulation of the target poses slightly better, especially when the arms are up.
The results illustrated in the main paper are generated using the first variation.

We think these distortions are related to the affine-invariance of MVC, and should be improved with Green Coordinates.

\setlength{\tabcolsep}{0pt}\scriptsize
\renewcommand{\arraystretch}{0.01}
\begin{longtable}{*{3}{m{0.11\linewidth}}:*{3}{m{0.11\linewidth}}:*{3}{m{0.11\linewidth}} }
\caption{The deformation results of the \textit{chair} category.}\label{tab:chairs}\\
	\makecell{Source}&\makecell{Deformed}&\makecell{Target}&\makecell{Source}&\makecell{Deformed}&\makecell{Target}&\makecell{Source}&\makecell{Deformed}&\makecell{Target}\\\midrule
	\endfirsthead
	\\
	\multicolumn{9}{l}{\textit{Continued from the last page}}\\\midrule
	{Source}&{Deformed}&{Target}&{Source}&{Deformed}&{Target}&{Source}&{Deformed}&{Target} \\\midrule
	\endhead
	\\\midrule\multicolumn{9}{l}{\textit{Continued on the next page}} \\
	\endfoot
	\endlastfoot
	\multicolumn{3}{p{0.33\linewidth}:}{\includegraphics[width=\linewidth]{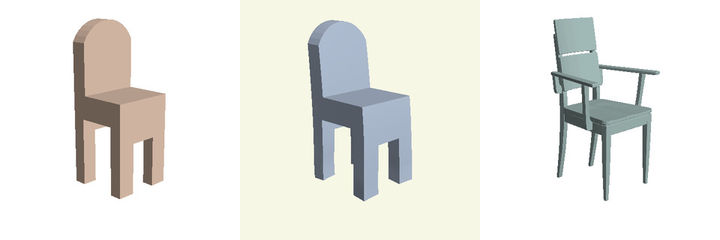}}&
	\multicolumn{3}{p{0.33\linewidth}:}{\includegraphics[width=\linewidth]{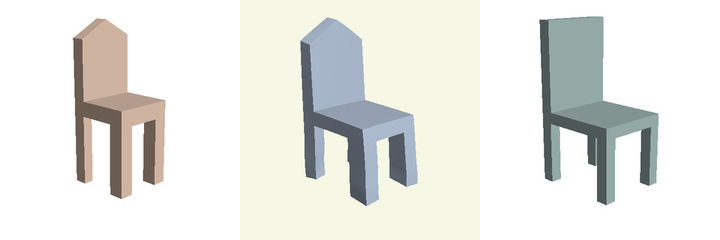}}&
	\multicolumn{3}{p{0.33\linewidth}}{\includegraphics[width=\linewidth]{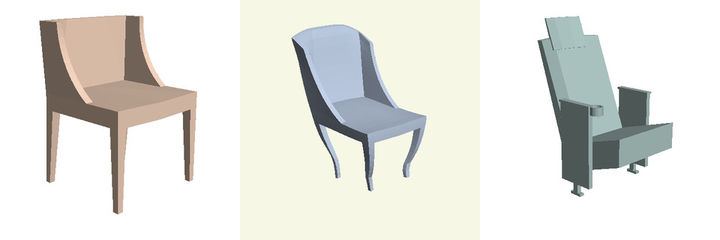}}\\
	\multicolumn{3}{p{0.33\linewidth}:}{\includegraphics[width=\linewidth]{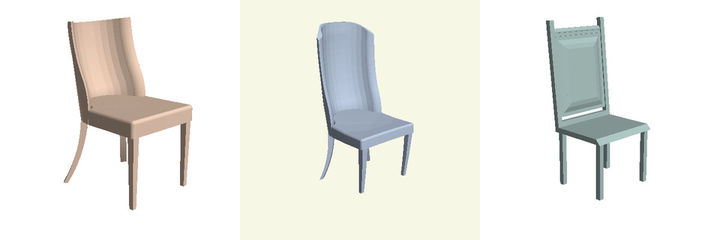}}&
	\multicolumn{3}{p{0.33\linewidth}:}{\includegraphics[width=\linewidth]{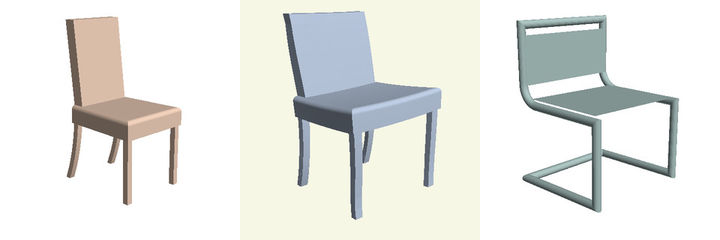}}&
	\multicolumn{3}{p{0.33\linewidth}}{\includegraphics[width=\linewidth]{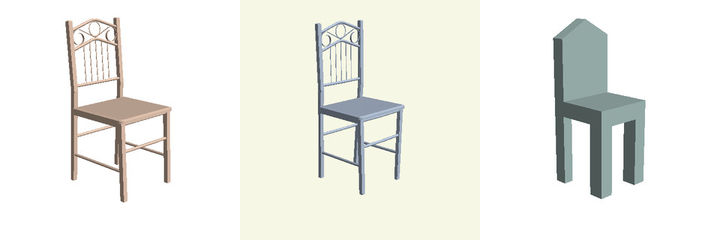}}\\
	\multicolumn{3}{p{0.33\linewidth}:}{\includegraphics[width=\linewidth]{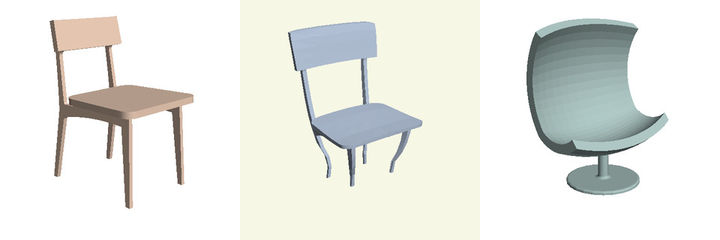}}&
	\multicolumn{3}{p{0.33\linewidth}:}{\includegraphics[width=\linewidth]{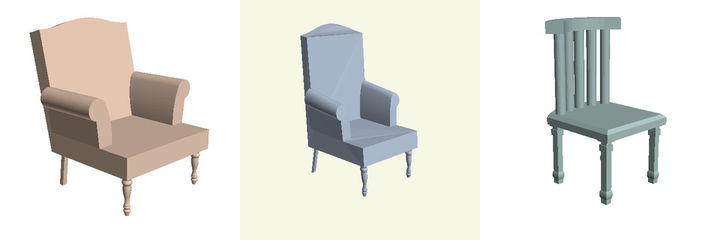}}&
	\multicolumn{3}{p{0.33\linewidth}}{\includegraphics[width=\linewidth]{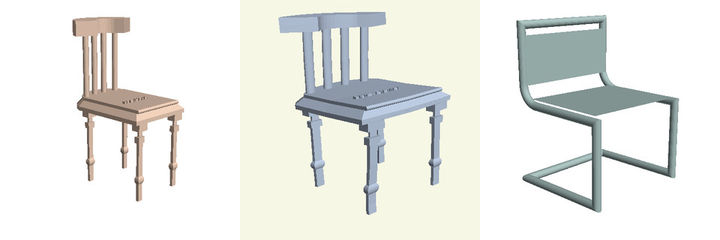}}\\
	\multicolumn{3}{p{0.33\linewidth}:}{\includegraphics[width=\linewidth]{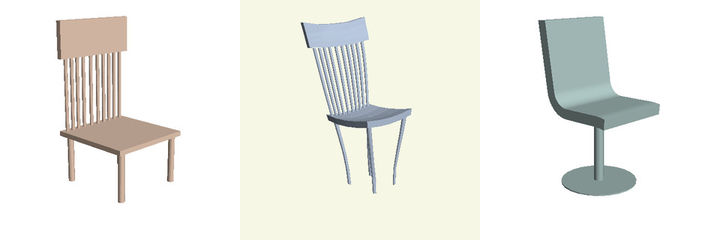}}&
	\multicolumn{3}{p{0.33\linewidth}:}{\includegraphics[width=\linewidth]{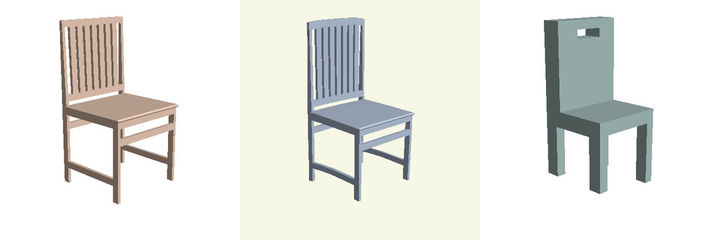}}&
	\multicolumn{3}{p{0.33\linewidth}}{\includegraphics[width=\linewidth]{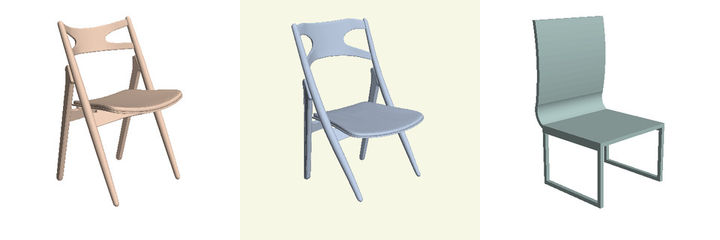}}\\
	\multicolumn{3}{p{0.33\linewidth}:}{\includegraphics[width=\linewidth]{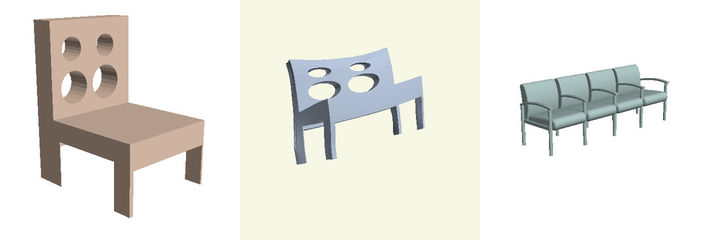}}&
	\multicolumn{3}{p{0.33\linewidth}:}{\includegraphics[width=\linewidth]{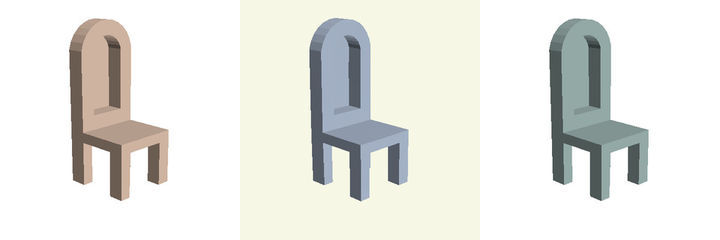}}&
	\multicolumn{3}{p{0.33\linewidth}}{\includegraphics[width=\linewidth]{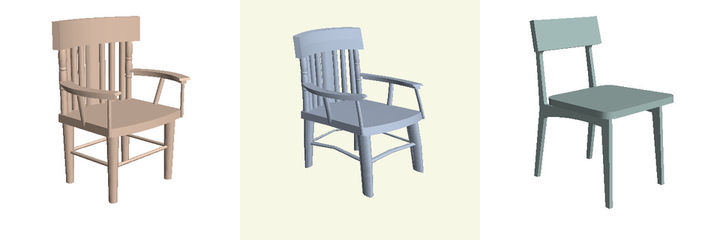}}\\
	\multicolumn{3}{p{0.33\linewidth}:}{\includegraphics[width=\linewidth]{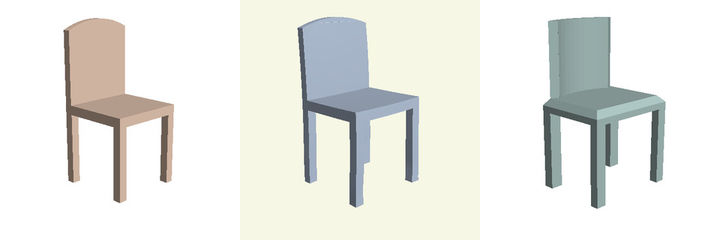}}&
	\multicolumn{3}{p{0.33\linewidth}:}{\includegraphics[width=\linewidth]{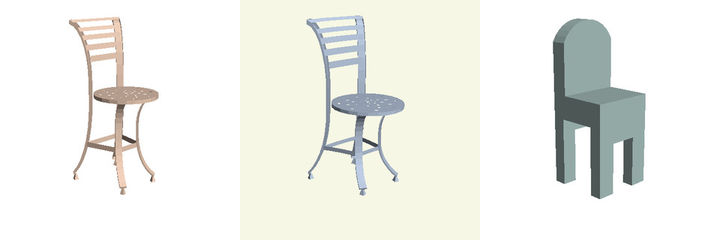}}&
	\multicolumn{3}{p{0.33\linewidth}}{\includegraphics[width=\linewidth]{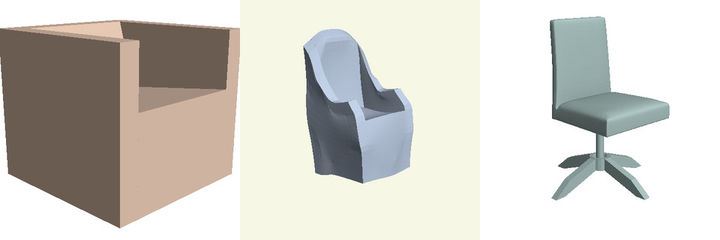}}\\
\bottomrule
\end{longtable}	

\begin{longtable}{*{3}{m{0.11\linewidth}}:*{3}{m{0.11\linewidth}}:*{3}{m{0.11\linewidth}} }
	\caption{The deformation results of the \textit{table} category.}	\label{tab:tables}\\
	\makecell{Source}&\makecell{Deformed}&\makecell{Target}&\makecell{Source}&\makecell{Deformed}&\makecell{Target}&\makecell{Source}&\makecell{Deformed}&\makecell{Target}\\\midrule
	\endfirsthead
	\\
	\multicolumn{9}{l}{\textit{Continued from the last page}}\\\midrule
	{Source}&{Deformed}&{Target}&{Source}&{Deformed}&{Target}&{Source}&{Deformed}&{Target} \\\midrule
	\endhead
	\\\midrule\multicolumn{9}{l}{\textit{Continued on the next page}} \\
	\endfoot
	\endlastfoot
	\multicolumn{3}{p{0.33\linewidth}:}{\includegraphics[width=\linewidth]{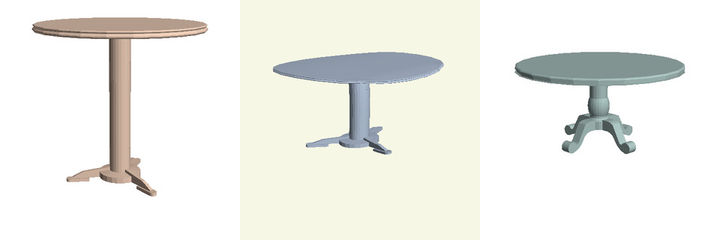}}&
	\multicolumn{3}{p{0.33\linewidth}:}{\includegraphics[width=\linewidth]{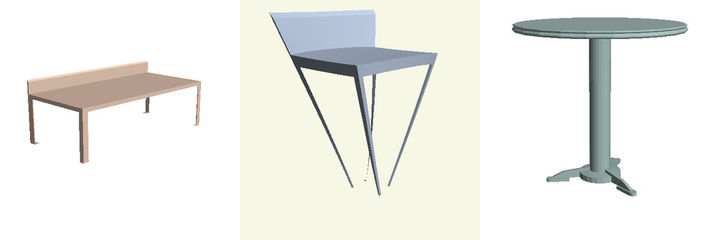}}&
	\multicolumn{3}{p{0.33\linewidth}}{\includegraphics[width=\linewidth]{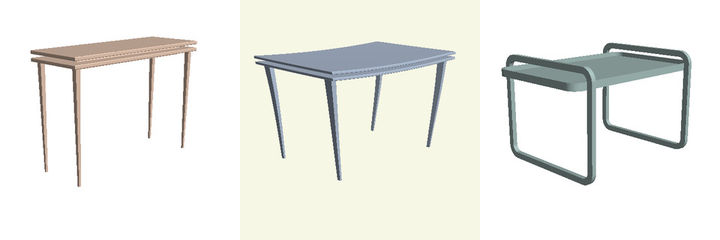}}\\
	\multicolumn{3}{p{0.33\linewidth}:}{\includegraphics[width=\linewidth]{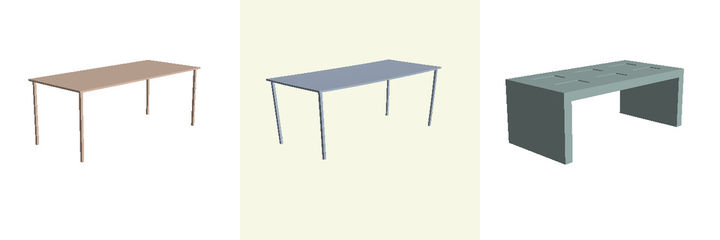}}&
	\multicolumn{3}{p{0.33\linewidth}:}{\includegraphics[width=\linewidth]{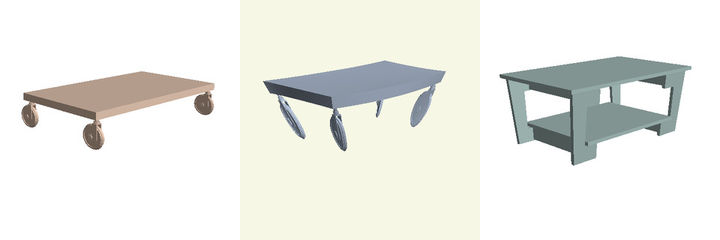}}&
	\multicolumn{3}{p{0.33\linewidth}}{\includegraphics[width=\linewidth]{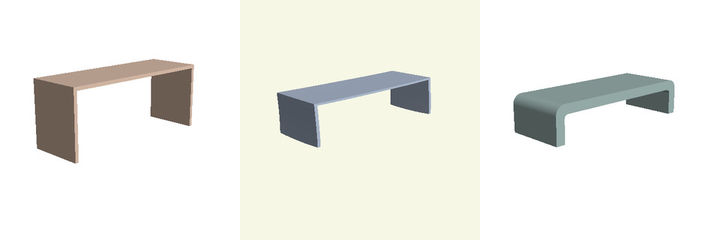}}\\
	\multicolumn{3}{p{0.33\linewidth}:}{\includegraphics[width=\linewidth]{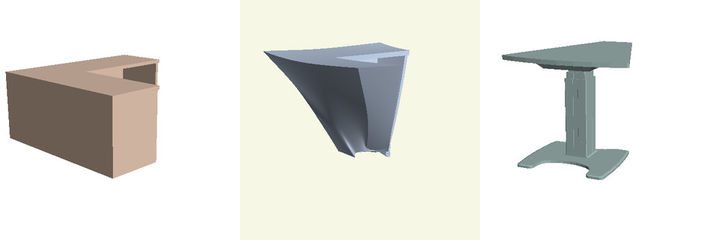}}&
	\multicolumn{3}{p{0.33\linewidth}:}{\includegraphics[width=\linewidth]{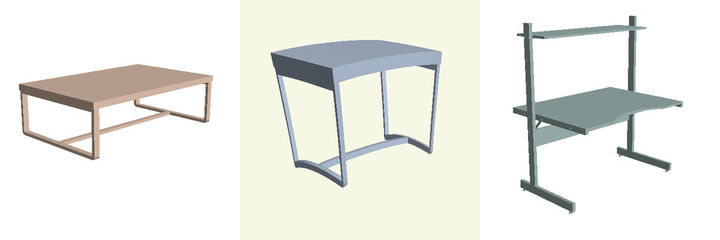}}&
	\multicolumn{3}{p{0.33\linewidth}}{\includegraphics[width=\linewidth]{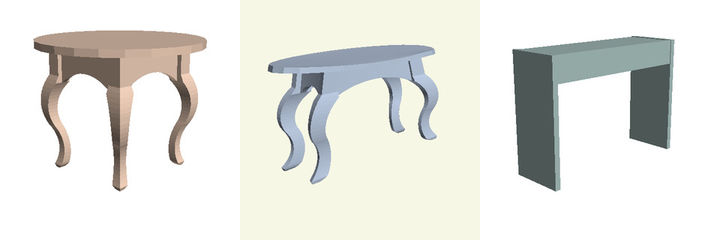}}\\
	\multicolumn{3}{p{0.33\linewidth}:}{\includegraphics[width=\linewidth]{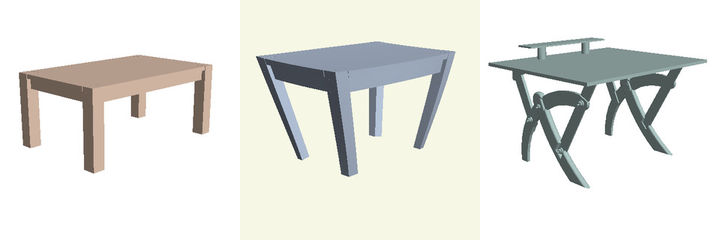}}&
	\multicolumn{3}{p{0.33\linewidth}:}{\includegraphics[width=\linewidth]{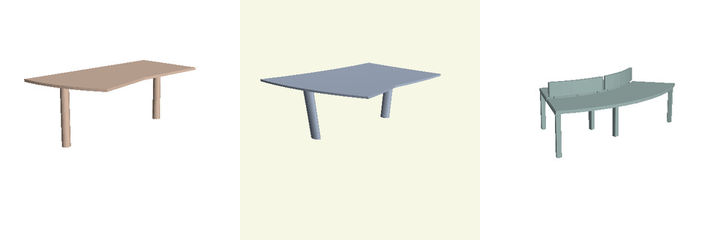}}&
	\multicolumn{3}{p{0.33\linewidth}}{\includegraphics[width=\linewidth]{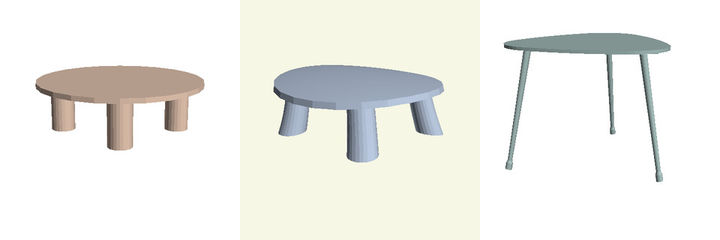}}\\
	\multicolumn{3}{p{0.33\linewidth}:}{\includegraphics[width=\linewidth]{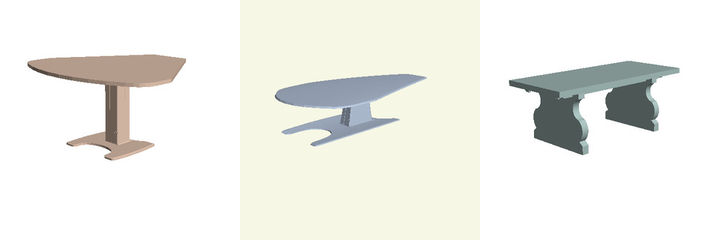}}&
	\multicolumn{3}{p{0.33\linewidth}:}{\includegraphics[width=\linewidth]{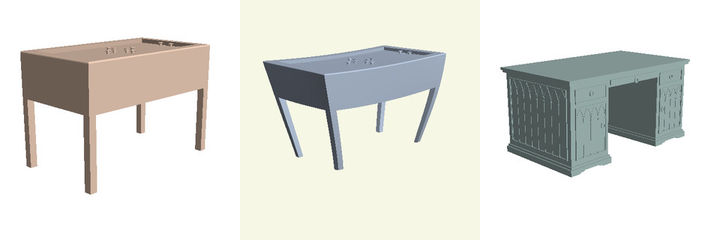}}&
	\multicolumn{3}{p{0.33\linewidth}}{\includegraphics[width=\linewidth]{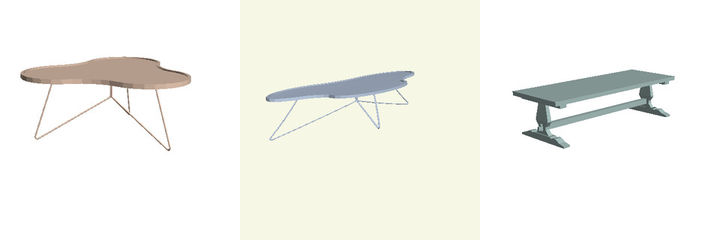}}\\
	\multicolumn{3}{p{0.33\linewidth}:}{\includegraphics[width=\linewidth]{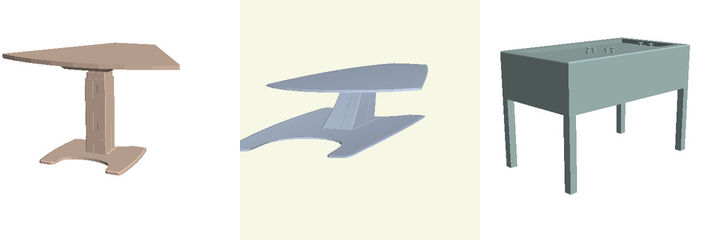}}&
	\multicolumn{3}{p{0.33\linewidth}:}{\includegraphics[width=\linewidth]{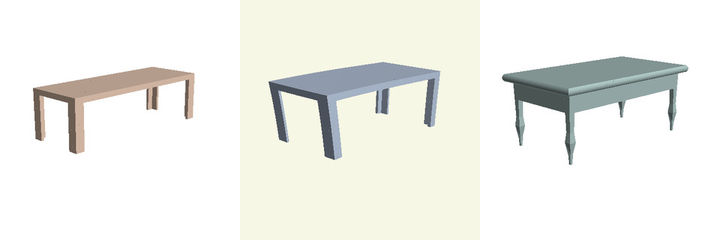}}&
	\multicolumn{3}{p{0.33\linewidth}}{\includegraphics[width=\linewidth]{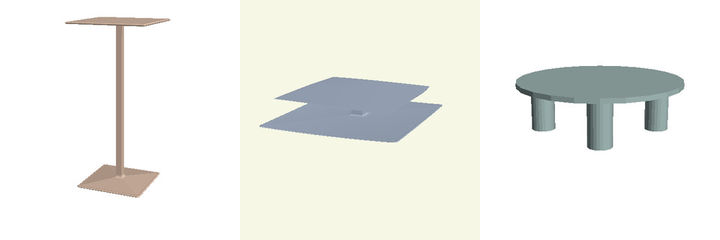}}\\
\bottomrule
\end{longtable}	

\begin{longtable}{*{3}{m{0.11\linewidth}}:*{3}{m{0.11\linewidth}}:*{3}{m{0.11\linewidth}} }	\caption{The deformation results of the \textit{car} category.}	\label{tab:cars}\\
	\makecell{Source}&\makecell{Deformed}&\makecell{Target}&\makecell{Source}&\makecell{Deformed}&\makecell{Target}&\makecell{Source}&\makecell{Deformed}&\makecell{Target}\\\midrule
	\endfirsthead
	\\
	\multicolumn{9}{l}{\textit{Continued from the last page}}\\\midrule
	\makecell{Source}&\makecell{Deformed}&\makecell{Target}&\makecell{Source}&\makecell{Deformed}&\makecell{Target}&\makecell{Source}&\makecell{Deformed}&\makecell{Target} \\\midrule
	\endhead
	\\\midrule\multicolumn{9}{l}{\textit{Continued on the next page}} \\
	\endfoot
	\endlastfoot
	\multicolumn{3}{p{0.33\linewidth}:}{\includegraphics[width=\linewidth]{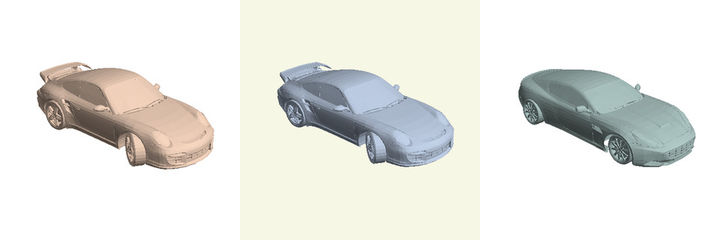}}&
	\multicolumn{3}{p{0.33\linewidth}:}{\includegraphics[width=\linewidth]{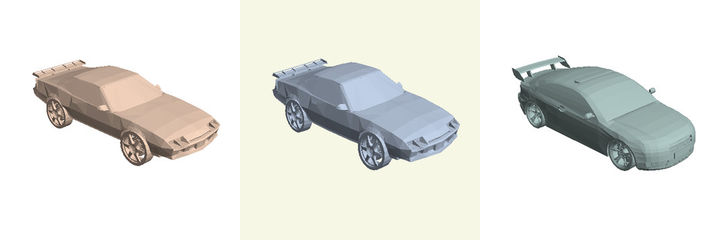}}&
	\multicolumn{3}{p{0.33\linewidth}}{\includegraphics[width=\linewidth]{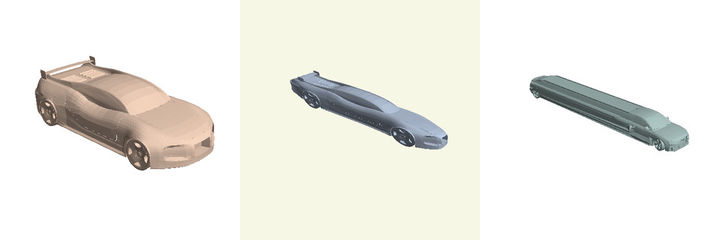}}\\
	\multicolumn{3}{p{0.33\linewidth}:}{\includegraphics[width=\linewidth]{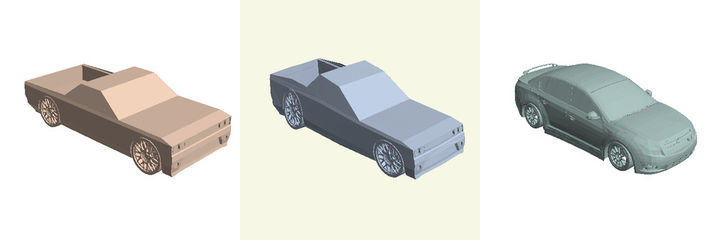}}&
	\multicolumn{3}{p{0.33\linewidth}:}{\includegraphics[width=\linewidth]{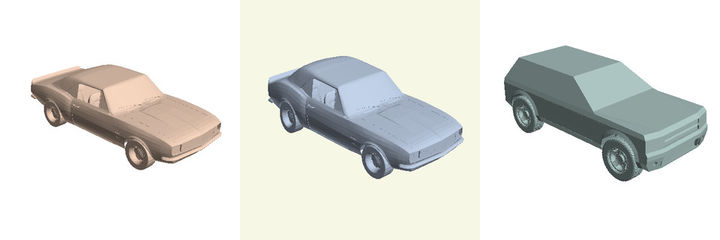}}&
	\multicolumn{3}{p{0.33\linewidth}}{\includegraphics[width=\linewidth]{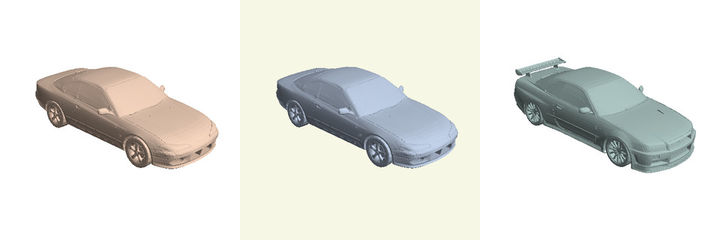}}\\
	\multicolumn{3}{p{0.33\linewidth}:}{\includegraphics[width=\linewidth]{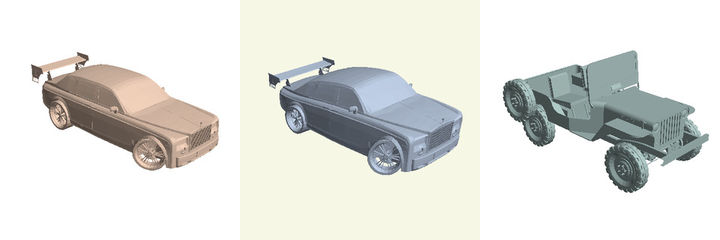}}&
	\multicolumn{3}{p{0.33\linewidth}:}{\includegraphics[width=\linewidth]{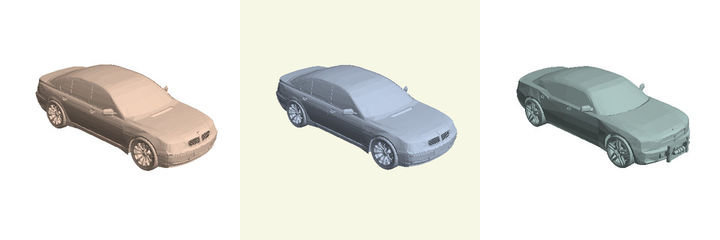}}&
	\multicolumn{3}{p{0.33\linewidth}}{\includegraphics[width=\linewidth]{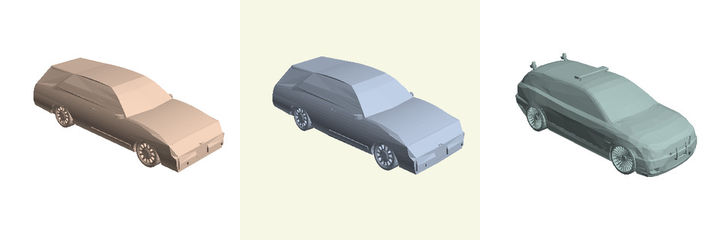}}\\
	\multicolumn{3}{p{0.33\linewidth}:}{\includegraphics[width=\linewidth]{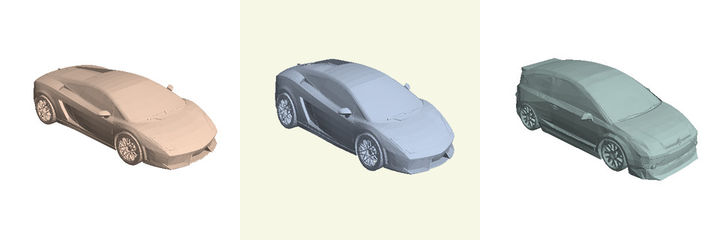}}&
	\multicolumn{3}{p{0.33\linewidth}:}{\includegraphics[width=\linewidth]{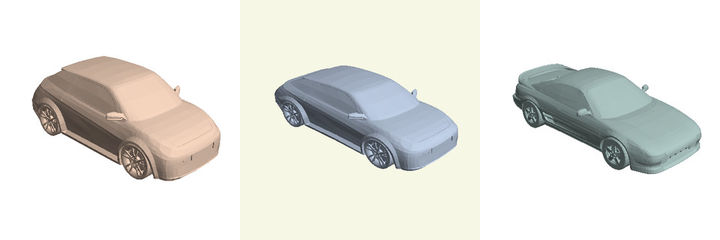}}&
	\multicolumn{3}{p{0.33\linewidth}}{\includegraphics[width=\linewidth]{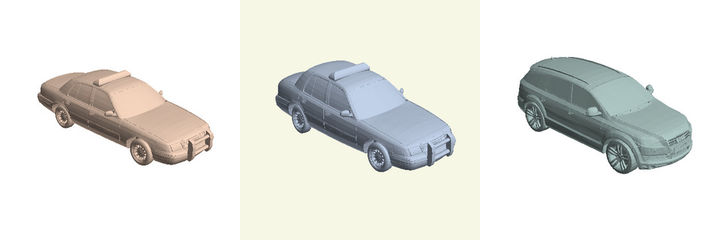}}\\
	\multicolumn{3}{p{0.33\linewidth}:}{\includegraphics[width=\linewidth]{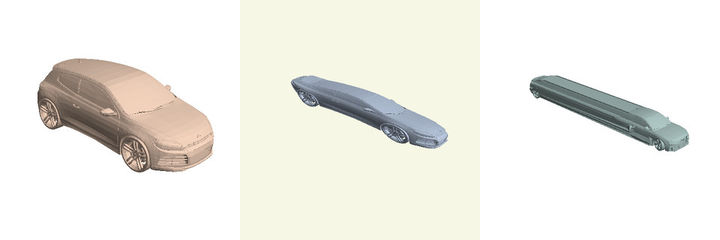}}&
	\multicolumn{3}{p{0.33\linewidth}:}{\includegraphics[width=\linewidth]{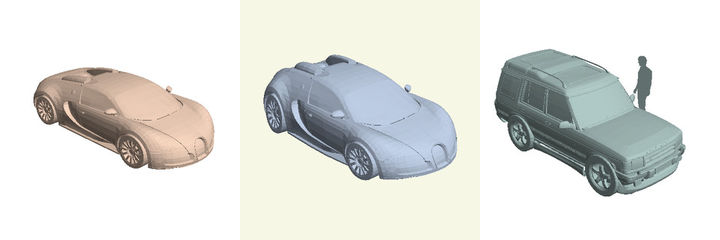}}&
	\multicolumn{3}{p{0.33\linewidth}}{\includegraphics[width=\linewidth]{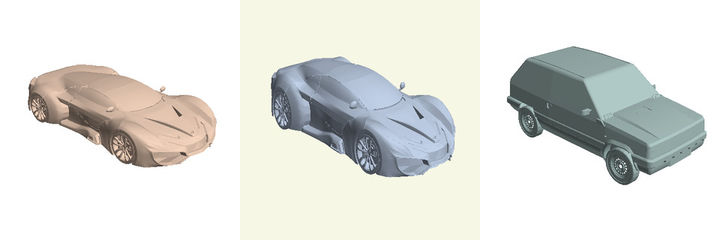}}\\
	\multicolumn{3}{p{0.33\linewidth}:}{\includegraphics[width=\linewidth]{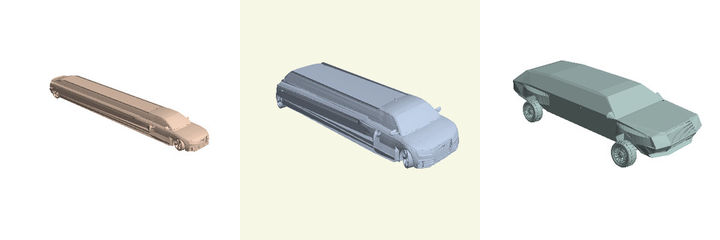}}&
	\multicolumn{3}{p{0.33\linewidth}:}{\includegraphics[width=\linewidth]{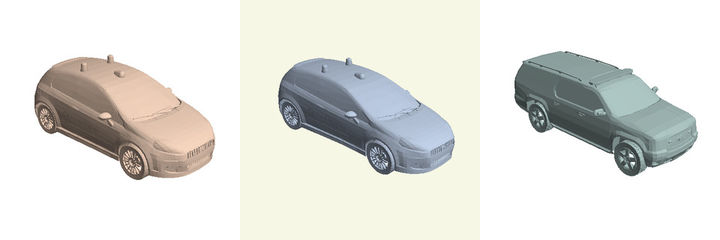}}&
	\multicolumn{3}{p{0.33\linewidth}}{\includegraphics[width=\linewidth]{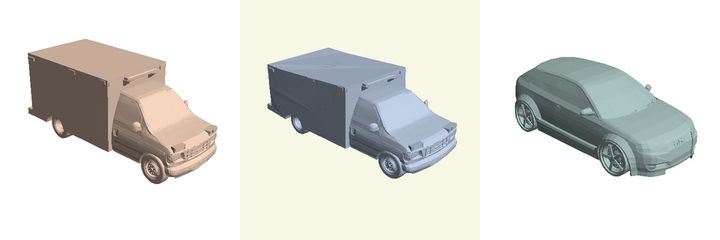}}\\
\bottomrule
\end{longtable}	

\begin{longtable}{p{0.11\linewidth}:*{4}{p{0.11\linewidth}}:*{4}{p{0.11\linewidth}}}
\caption{Additional results for humanoid deformation and deformation transfer to new characters. The target poses are shown on the left in green, the template source (fixed during training) and the novel new sources are shown on the top in brown. We exhibit two training results, using a rest pose template source and using a t-pose template source, shown on the left and right half of the table respectively.}\label{tab:dt}\\
\makecell{target} & \makecell{training source} &\multicolumn{3}{m{0.33\linewidth}}{\makecell{novel sources}} & \makecell{training source} & \multicolumn{3}{m{0.33\linewidth}}{\makecell{novel source}}\\
&{\includegraphics[width=\linewidth]{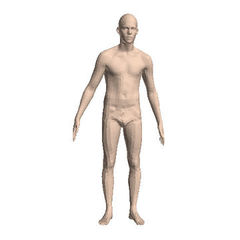}} &
{\includegraphics[width=\linewidth]{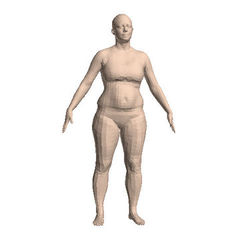}} &
{\includegraphics[width=\linewidth]{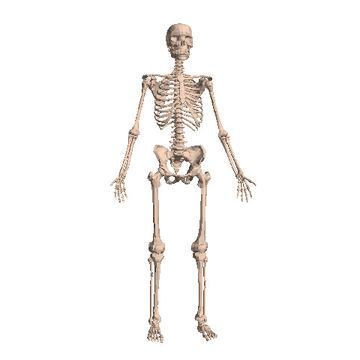}} &
{\includegraphics[width=\linewidth]{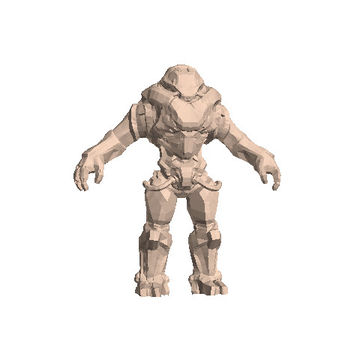}} &
{\includegraphics[width=\linewidth]{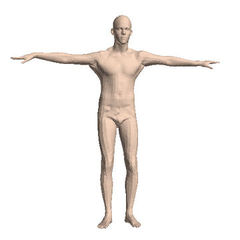}} &
{\includegraphics[width=\linewidth]{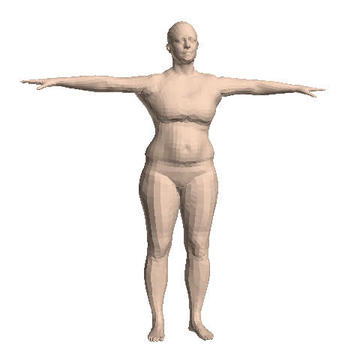}} &
{\includegraphics[width=\linewidth]{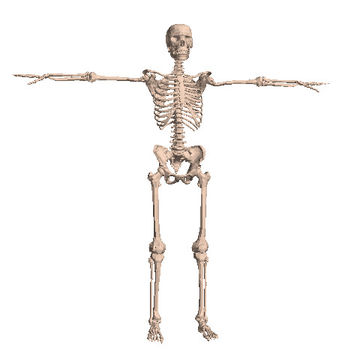}} &
{\includegraphics[width=\linewidth]{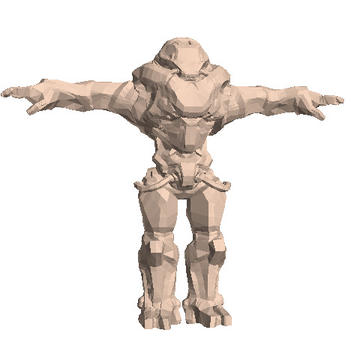}} \\\midrule
\endfirsthead
%\\
\multicolumn{9}{l}{\textit{Continued from the last page}}\\\midrule
\makecell{target} & \makecell{training source} &\multicolumn{3}{m{0.33\linewidth}}{\makecell{novel sources}} & \makecell{training source} & \multicolumn{3}{m{0.33\linewidth}}{\makecell{novel source}}\\
&\makecell{\includegraphics[width=\linewidth]{figures/supp_dt/my_surreal_test/template-0000-Sa}} &
\makecell{\includegraphics[width=\linewidth]{figures/supp_dt/my_surreal_retarget/template-Sa}} &
\makecell{\includegraphics[width=\linewidth]{figures/supp_dt/rest_opt_skeleton/Skeleton_normalized}} &
\makecell{\includegraphics[width=\linewidth]{figures/supp_dt/my_surreal_robot/robot_normalized}} &
\makecell{\includegraphics[width=\linewidth]{figures/supp_dt/tpose_test/template-0000-Sa}} &
\makecell{\includegraphics[width=\linewidth]{figures/supp_dt/tpose_retarget/faust_fat_female_normalized}} &
\makecell{\includegraphics[width=\linewidth]{figures/supp_dt/tpose_skeleton/skeleton_tpose_normalized}} &
\makecell{\includegraphics[width=\linewidth]{figures/supp_dt/tpose_robot/robot_tpose_normalized}} \\\midrule
\endhead
\\\midrule\multicolumn{7}{l}{\textit{Continued on the next page}} \\
\endfoot
\endlastfoot
\includegraphics[width=\linewidth]{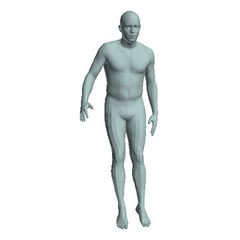}&
\includegraphics[width=\linewidth]{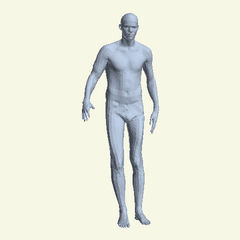}&
\includegraphics[width=\linewidth]{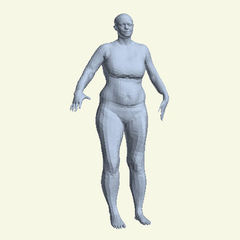}&
\includegraphics[width=\linewidth]{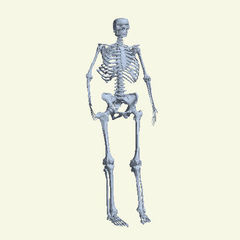}&
\includegraphics[width=\linewidth]{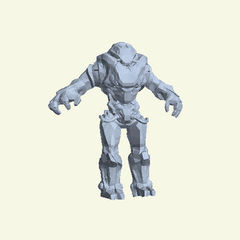}&
\includegraphics[width=\linewidth]{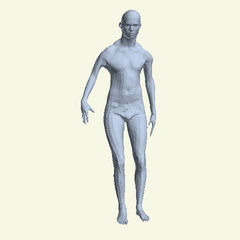} &
\includegraphics[width=\linewidth]{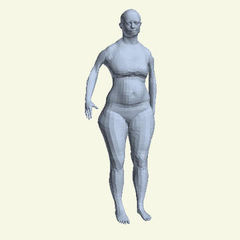} &
\includegraphics[width=\linewidth]{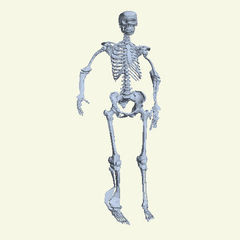} &
\includegraphics[width=\linewidth]{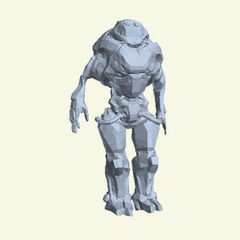} \\
\includegraphics[width=\linewidth]{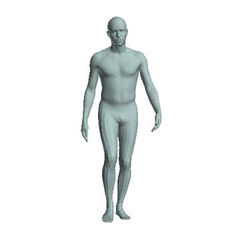}&
\includegraphics[width=\linewidth]{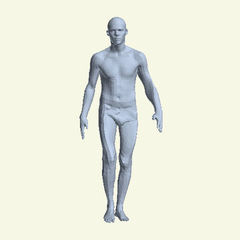}&
\includegraphics[width=\linewidth]{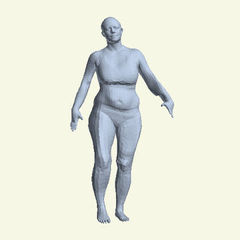}&
\includegraphics[width=\linewidth]{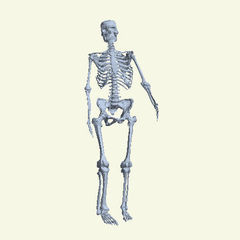}&
\includegraphics[width=\linewidth]{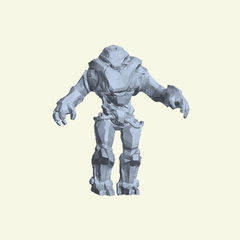}&
\includegraphics[width=\linewidth]{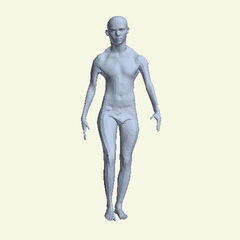} &
\includegraphics[width=\linewidth]{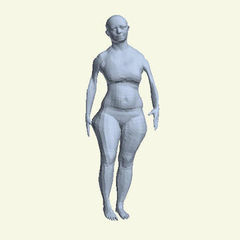} &
\includegraphics[width=\linewidth]{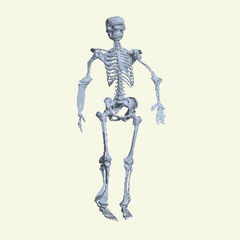} &
\includegraphics[width=\linewidth]{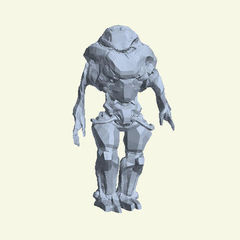}\\
\bottomrule
\end{longtable}

%{\small
%\bibliographystyle{ieee_fullname}
%\bibliography{egbib}
%}

%\end{document}